\documentclass[10pt,aps,prd,floatfix]{revtex4-1}

\usepackage{graphicx}
\usepackage{color}
\usepackage{amsmath, amssymb}

\newcommand{\PP}{\mathbb{P}}

\newcommand{\beq}{\begin{equation}}
\newcommand{\eeq}{\end{equation}}
\newcommand{\beqa}{\begin{eqnarray}}
\newcommand{\eeqa}{\end{eqnarray}}

\renewcommand{\Re}{ \operatorname{Re}}
\renewcommand{\Im}{ \operatorname{Im}}

\begin{document}

\title{Comparison of Domain Wall Fermion Multigrid Methods}

\author{Peter Boyle}
\affiliation{HET Physics Department, Brookhaven National Laboratory, Upton, NY 11973, USA.}
\affiliation{School of Physics and Astronomy, University of Edinburgh, Edinburgh EH9 3JZ, UK}

\author{Azusa Yamaguchi}
\affiliation{School of Physics and Astronomy, University of Edinburgh, Edinburgh EH9 3JZ, UK}

\begin{abstract}
We present a detailed comparison of several recent and new approaches to multigrid solver algorithms suitable for the 
solution of 5d chiral fermion actions such as Domain Wall fermions in the Shamir formulation, and also
for the Partial Fraction and Continued Fraction overlap. 
Our focus is on the acceleration of gauge configuration sampling, and a
compact nearest-neighbour stencil is required to limit the calculational cost of obtaining a coarse operator.
This necessitates the coarsening of a nearest neighbour operator
to preserve sparsity in coarsened grids, unlike HDCG\cite{Boyle:2014rwa}.
We compare the approaches of references\cite{Yamaguchi:2016kop,Brower:2020xmc} and also several
new hybrid schemes. In this work we
introduce a new recursive Chebyshev polynomial based multigrid setup scheme. We
find that the approach of reference\cite{Yamaguchi:2016kop}, 
can both setup and then solve twice standard Shamir domain wall fermions faster than a single
solve with red-black preconditioned Conjugate Gradients\cite{CGNR} on large volumes near the physical up/down quark masses
and for modern GPU systems such as the Summit supercomputer. This is promising for the acceleration of HMC,
particularly if setup costs are shared across multiple Hasenbusch determinant factors. The setup scheme
is likely generally applicable to other fermion actions.
\end{abstract}

\maketitle

\section{Introduction}

Despite the development of revolutionary new multilevel solver algorithms for Wilson fermions
\cite{Luscher:2007se,Brannick:2007ue,Brannick:2007cc,Clark:2008nh,Babich:2009pc},
the extension of the approaches to all fermion actions remains somewhat piecemeal.
The generalisation to improved Wilson (clover) fermions was made rather rapidly\cite{Osborn:2010mb},
and subsequent variations \cite{Frommer:2012mv,Frommer:2013fsa,Frommer:2013kla} have included
more efficient subspace setup.
Multigrid algorithms for staggered fermions have quite recently been developed \cite{Weinberg:2017zlv,Brower:2018ymy}.
The extension to domain wall fermions\cite{Kaplan:1992bt,Shamir:1993zy} has been studied\cite{Cohen:2012sh} and an approach
made to give a substantial acceleration for valence analysis 
based on the red-black preconditioned squared operator\cite{Boyle:2014rwa}.
The stencil for the squared operator contains all points with taxicab norm less than four, giving
321 points in four dimensions. This has the result that approach is unattractive for 
gauge evolution code where, even \emph{if} the subspace quality can be
preserved along an HMC trajectory, the re-evaluation of the matrix elements of the little Dirac operator
on each time step in the integrator, for O(50) vectors in the subspace requires naively 15000 matrix multiplies.
Even admitting a constraint, such as a minimum block size of $4^4$, the squared operator stencil only reduces
to 81 points\cite{Boyle:2014rwa} and recalculation costs over 4000 matrix multiplies.

In this paper, we compare two approaches. That of the authors \cite{Yamaguchi:2016kop} and that of reference \cite{Brower:2020xmc},
which was introduced for the 2D Schwinger model.
We extend the latter to $D=4$ and $SU(3)$ gauge theory for the first time. These approaches enable a nearest neighbour
coarsening in two different ways, by coarsening the hermitian domain wall operator in the former and the Wilson operator
in the latter. They have in common the property that,
$
\Gamma_5 = \gamma_5 R_5,
$
commutes with the prolongation and restriction operations.

The structure of this paper is as follows:
We discuss the actions to which our methods apply in section~\ref{sec:actions}.
We discuss the general multigrid framework in section~\ref{sec:mgrid}, and 
introduce a novel approach to generating the subspace for coarse degrees of freedom in section~\ref{sec:chebyshev}.
We believe this approach is generally applicable and could be adopted for other actions.
We introduce the specific algorithms tested in this paper in section~\ref{sec:algorithms}.
Results for numerical efficiency are presented in section~\ref{sec:numerics} using $16^3\times 32$ and
$48^3\times 96$ test volumes, and our conclusions are drawn in section~\ref{sec:conclusions}.

\section{Chiral Fermion Actions}
\label{sec:actions}
We consider two classes of approach for chiral fermions following the nomenclature of ref. \cite{Kennedy:2006ax}.
We restrict our consideration to five dimensional approaches to chiral fermions since we view the non-locality associated with
nested four dimensional approaches chiral fermions as an avoidable difficulty for multigrid implementation.
As discussed in the next section, this decision is not without associated problems as it does introduce other difficulties
associated with the spectrum of the five dimensional operators.
Usual Wilson matrix is,
\beq D_W(M) = M+4 - \frac{1}{2} D_{\rm hop},\eeq 
where,
\begin{equation}
D_{\rm hop} = (1-\gamma_\mu) U_\mu(x) \delta_{x+\mu,y} +
              (1+\gamma_\mu) U_\mu^\dagger(y) \delta_{x-\mu,y}.
\end{equation}

The domain wall fermion action is,
\begin{equation}
S^5 = \int d^4x \bar{\psi} D^5_{DW} \psi,
\end{equation}
where,
\begin{equation}
D^5_{DW} = 
\left(
\begin{array}{cccccc}
D_\parallel & - P_-  &  0      & \ldots & 0 & m P_+ \\
-P_+  & \ddots &  \ddots & 0      & \ldots &0 \\
0     & \ddots &  \ddots & \ddots & 0      &\vdots \\
\vdots& 0      &  \ddots & \ddots & \ddots & 0\\
0     & \ldots &    0    &  \ddots& \ddots & -P_- \\
m P_- & 0      & \ldots  &  0     & -P_+   & D_\parallel
\end{array}
\right),
\end{equation}
and,
\beq
D_{\parallel} = 5 - M_5 - \frac{1}{2} D_{\rm hop} = D_W(-M_5)+1 .
\eeq
We introduce,
$
\Gamma_5 = \gamma_5 R_5,
$
where $R_5$ denote reflection in the fifth dimension. We define the hermitian indefinite DWF operator to be,
\beq
H_{DW} = \Gamma_5 D_{DW}.
\eeq
The Pauli Villars operator $D_{PV}$ is equal to $D_{DW}$ with unit mass parameter.

The continued fraction and partial fraction five dimensional representations
of the overlap operator for the
standard overlap $H_W=\gamma_5 D_W$ kernel are already hermitian indefinite\cite{Kennedy:2006ax},
with the continued fraction system taking the form,
\beq
\left[
\begin{array}{ccccc}
H_W & \frac{1}{\sqrt{\beta_0\beta_1}} & 0 & 0 &0 \\
\frac{1}{\sqrt{\beta_1\beta_0}}   & -H_W  & \frac{1}{\sqrt{\beta_1\beta_2}}   & 0 & 0  \\
0   & \frac{1}{\sqrt{\beta_2\beta_1}}   & H_W   & \frac{1}{\sqrt{\beta_2\beta_3}} & 0  \\
0   & 0   & \frac{1}{\sqrt{\beta_3\beta_2}}   & -H_W & \frac{1}{\sqrt{\beta_3}} \\
0   & 0   & 0   & \frac{1}{\sqrt{\beta_3}} & R\gamma_5 + \beta_0 H_W
\end{array}
\right],
\eeq
and the partial fraction system taking the form,
\beq
\left[ 
\begin{array}{cc|cc|c}
-H_W & \sqrt{q_2}         & 0 &0 & 0\\
\sqrt{q_2}        & H_W   & 0 &0 & \sqrt{p_2}\\
\hline
 0 &0 & -H_W & \sqrt{q_1} & 0 \\
 0 &0 & \sqrt{q_1} & H_W  & \sqrt{p_1}\\
\hline
0 &  -\sqrt{p_2} &0 & -\sqrt{p_1}  & R\gamma_5 + p_0 H_W
\end{array}
\right].
\eeq
These are both amenable to all the multigrid methods discussed in this paper,
and we have demonstrated that the continued fraction approach may be solved with a two level HDCR
algorithm. However we will not present results as the focus is on the domain wall and Mobius fermion
actions. The convergence has been tested numerically in the continued fraction case by the authors.

\subsection{Spectrum of domain wall fermions}
\label{sec:spectrum}
It is clear that in order to make a practical algorithm for accelerating
HMC evolution with domain wall fermions we must escape the constraint that the algorithm work on the squared
operator (or worse the squared red-black preconditioned operator). 
In order to do this we must first understand why prior to reference\cite{Yamaguchi:2016kop} only solvers making use of the squared
operator have been successful for domain wall fermions.

\begin{figure}[hbt]
\includegraphics*[width=\textwidth]{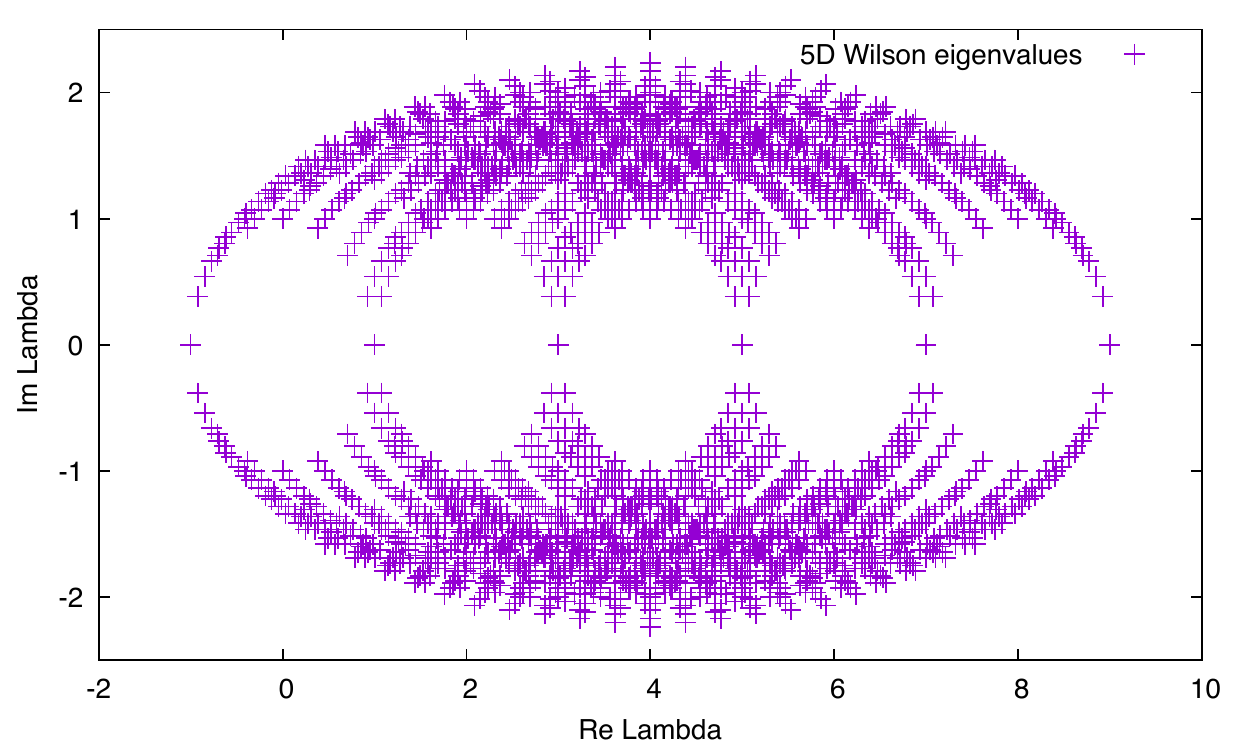}
\caption{\label{fig:hamburger}
The complex eigenvalue spectrum of the 5D Wilson operator on a $16^5$ free field.
With a negative Wilson mass of order 1.0, as is introduced in the kernel of chiral fermions, the spectrum completely encircles zero.
In the infinite volume limit this spectrum becomes continuous, and poses clear problems for polynomial based Krylov solvers.
}
\end{figure}

The spectrum of the free 5d Wilson operator at negative mass is illustrated in figure~\ref{fig:hamburger}.
The eigenvalues have $\Re \lambda = m + 5 - \sum\limits \cos p_\mu $ and $|\Im \lambda|^2 = \sum\limits \sin^2 p_\mu $.
It is a reasonable illustrative guide for that of domain wall fermion operator, differing only by the free field
approximation and the fifth dimension boundary condition. Complete analysis of the DWF propagator including the
Dirichlet boundary conditions and surface states is given in reference\cite{Aoki:1997xg}.

The spectrum for an appropriate negative mass completely encircles the origin and violates the {\emph{half-plane condition}}
referred to in numerical analysis literature\cite{trefethen}. There is a fundamental reason for this:
in the infinite volume the spectrum will become dense, and the Krylov solver is then being asked to form
an (analytic) polynomial approximation to $\frac{1}{z}$ over an open region encircling the pole, and the
Cauchy residue theorem will apply to the error. It is impossible
to reproduce the phase winding of $\frac{1}{z}=\frac{1}{r}e^{-i\theta}$
around zero with an analytic function.  Indeed, perhaps belabouring the point,
the orthogonality of the set of functions $e^{i m\theta}$ over $[0,2\pi]$ makes
it easy to show that minimising the uniformly weighted mean square error over a
fixed radius circle gives precisely zero for all polynomial coefficients. 
In the case of Conjugate Gradient on the Normal Residual  (CGNR),
the multiplication of each eigenvalue by its conjugate in solving,
\beq M_{pc}^\dag M_{pc} \psi =  \eta, \eeq
places the phase behaviour under control and reduces the problem to a real spectrum, albeit with a squared range of eigenvalue magnitudes. CGNR is used to date in RBC-UKQCD domain wall fermion evolution.
There is, in principle, a reduced convergence rate arising from the squared
changed condition number in the convergence bound\cite{Saad},
\beq
\sigma = \frac{\sqrt{K} -1 }{\sqrt{K}+1},
\eeq
where $\sigma$ is the residual reduction in one iteration, and $K=\frac{\lambda_{\mathrm{max}}}{\lambda_{\mathrm{min}}}$ is the condition number of the
matrix. On the free field on the unpreconditioned normal equations the maximal eigenvalue is of O(100).

In the discrete spectrum, finite volume case, we can consider a toy models which also illustrate the problem. If the spectrum consists
of $N$ eigenvalues $\lambda_k = e^{i2\pi k/N}$ the conjugate gradient will only converge with an N-term polynomial, which can be analytically arrived
at by Gaussian elimination for small $N$\cite{trefethen}.

In reference\cite{Yamaguchi:2016kop}, the authors proposed to solve this
phase problem using $\Gamma_5$ hermiticity, without squaring the operator,
leaving the coarse space representation of the operator still nearest neighbour. Since the sparsity pattern is preserved
this represented the first true multigrid algorithm for five dimensional chiral fermions.

With a restrictor that is $\Gamma_5$ compatible, HDCR can be thought of as either solving a hermitian indefinite
coarsened system, or the squared coarse operator in a normal equations sense. 
\beq
\PP^\dagger \Gamma_5 D_{DW} \PP \PP^\dagger \Gamma_5 D_{DW} \PP = \PP^\dagger D_{DW}^\dagger \PP \PP^\dagger D_{DW} \PP
\eeq
$\gamma_5$ compatible coarsenings have been used for some time in Wilson multigrid\cite{Frommer:2013fsa}.
While well understood, it is probably worth some comments on the advantages.
$D_W$ is a non-hermitian, non-normal operator. Its left and right eigenvectors do not coincide and its singular
value decomposition (SVD) takes the form
$
D_{DW} = U D V^\dagger
$, while a general normal matrix has left and right eigenvectors coincide and has SVD
$
V D V^\dagger
$.
A hermitian matrix takes the same form as a normal matrix with real eigenvalues contained in $D$.
Preservation of $\gamma_5$ hermiticity in coarsening was initially debated in the development of multigrid
for Wilson and clover fermions \cite{Luscher:2007se,Babich:2010qb,Osborn:2010mb}.
We believe that a key point is that the $\gamma_5$ hermitian
operator is normal, and guarantees that the left and right null spaces coincide.
A $\gamma_5$ compatible coarsening of $D_W$ is equivalent in span to a coarsening of $H_W$.

These $\Gamma_5$ hermitian operator is \emph{nearest neighbour} in the space-time dimensions
and preserves sparsity in a coarse space with a four dimensional coarsening. The hermiticity gives rise to a
real \emph{indefinite} spectrum, and the squared eigenvalues are the spectrum of the hermitian positive
definite squared operator. As a theoretical exercise, in the infinite volume the 
spectrum will be dense, real and symmetrical about the origin. From the perspective of a Krylov solver the
polynomial approximation $P(\lambda)\sim \frac{1}{\lambda}$ must be made over a the subset of
real line $\lambda\in [-\lambda_{\rm max},-\lambda_{\rm min}] \cup [\lambda_{\rm min},\lambda_{\rm max}] $

Such a spectrum succumbs easily to the (generalised) conjugate
residual algorithm, which relaxes the hermitian positive definite
constraint of conjugate gradients to only hermitian indefinite.
We therefore use variants of conjugate residuals as the basis of the  outer fine matrix iteration. 
Regarding the relative efficiency, it is worth to note that we create a Krylov space 
that strictly contains the CGNE Krylov space (spanned by every second term),
\beq
P_N( D^\dag D ) D^\dag = P_N( \Gamma_5 D \Gamma_5 D ) \gamma_5 D \gamma_5
\subset P_{2N+1}( H_{DW} ) \gamma_5 =  P_{2N+1}( \Gamma_5 D ) \gamma_5 .
\eeq
Further, since either on average or in the infinite volume, the spectrum will be symmetrical about zero,
the even terms cannot contribute to an approximation of the (odd) function $\frac{1}{x}$ and the in this limit the 
iteration should converge with an identical number of applications of the 
nearest neighbour fermion operator as unpreconditioned CGNE. This limit is observed to be practically true even on $16^3$
configurations.

\subsubsection{Pauli-Villars preconditioning}
  
It was shown in refererence\cite{Brower:2020xmc} that the fine operator, $$M_{PV}^\dagger M_l,$$ satisfies the half
plane condition and represents a wholly new alternative for preconditioning domain wall fermions. It potentially raises
the range of the required coarse space Krylov polynomial from $[m_l^2,64]$ to $[m_l,64]$, at the expense of
treating a non-hermitian squared operator. Finding evidence of algorithmic
benefit from this direction is one goal of this work, as it will indicate whether further gains may be possible
using this direction. The matrix is projected to the coarse space and approximated as, 
\beq
\PP^\dagger D_{PV}^\dagger \PP \PP^\dagger D_{DW} \PP.
\eeq
In this work we study this system in the case of SU(3) gauge theory and four dimensions for the first time.
In principle, with a non-hermitian solver,
one can tune the adjoint matrix mass continuously between the Pauli Villars mass and the
light quark mass between the usual squared operator limits and this new Pauli Villars preconditioning idea.
The spectrum will vary between being real with lowest eigenvalue of $O(m_l^2)$, and complex with a larger minimal
real component.

\section{Multigrid algorithms}

The fundamental composable element of multigrid algorithms is the two level preconditioner. These
may then be nested recursively whenever the sparsity pattern is preserved in the coarsening.
These are combined with smoothers, and introduced as a multigrid correction step to Krylov process as a preconditioner,
we use either variable preconditioned GCR\cite{GCR}, BiCGSTAB\cite{BiCGSTAB} or CG\cite{CGNR} as the outer iteration.
These Krylov solvers are standard algorithms that we will not document in the interests of brevity, and in this section document
the elements of a multigrid preconditioner used in this work. These are a two level preconditioner, smoothers, and
subspace generation.

\subsection{Two level preconditioner}
\label{sec:mgrid}

The key element is the selection of a \emph{deflation basis} of vectors $\phi_k$ that lie in the near null space of the Dirac operator. The details of how these are selected
are immaterial for the present discussion, but we will return to this in the following section.  The details affect setup cost, and deflation efficiency.
The vectors $\phi_k$ are then restricted to blocks, enabling a
coarse space representation to be built up as follows,
\beq
\phi^b_k(x) = \left\{ \begin{array}{ccc}
  \phi_k(x) &;& x\in b\\
  0 &;& x \not\in b
\end{array}
\right. .
\eeq
The span of these blocks is substantially larger than the span of the originial vector set,
\beq{\rm span} \{ \phi_k\}\subset
{\rm span} \{ \phi_k^b\} .\eeq
and it is by
now well demonstrated to capture the near null space of the operator. This property is variously known as the \emph{weak approximation} property \cite{Brezina},
\emph{local coherence}\cite{Luscher:2007se} and has been demonstrated as effective for data compression of individual eigenvectors\cite{Clark:2017wom}.

The fine operator is used to generate a coarse space representation that faithfully represents the matrix on this subspace, and the inverse of the
subspace restriction is used to accelerate convergence since by design this subspace encapsulates modes with the smallest eigenvalues that are the slowest
to converge in a Krylov solver.
We introduce projectors to the subspace $S$ and its complement $\bar{S}$,
\beq
P_S =  \sum_{k,b} |\phi^b_k\rangle \langle \phi^b_k | \quad\quad ; \quad\quad P_{\bar{S}} = 1 - P_S,
\eeq
and can decompose the matrix into terms within and between $S$ and $\bar{S}$,
\beq
M=
\left(
\begin{array}{cc}
M_{\bar{S}\bar{S}} & M_{S\bar{S}}\\
M_{\bar{S}S} &M_{SS}
\end{array}
\right)=
\left(
\begin{array}{cc}
P_{\bar{S}} M P_{\bar{S}}  &  P_S M P_{\bar{S}}\\
 P_{\bar{S}} M P_S &   P_S M P_S
\end{array}
\right).
\eeq
We can represent the matrix $M$ exactly on this subspace by computing its matrix elements,
known as the \emph{little Dirac operator} (coarse grid matrix in multi-grid),
\beq
A^{ab}_{jk} = \langle \phi^a_j| M | \phi^b_k\rangle
\quad\quad ; \quad\quad
(M_{SS}) = A_{ij}^{ab} |\phi_i^a\rangle \langle \phi_j^b |.
\eeq
The subspace inverse can be solved by Krylov methods and is,
\beq
Q =
\left( \begin{array}{cc}
0 & 0 \\ 0 & M_{SS}^{-1}
\end{array} \right)
\quad\quad ; \quad\quad
M_{SS}^{-1} = (A^{-1})^{ab}_{ij} |\phi^a_i\rangle \langle \phi^b_j |.
\eeq
It is important to note that $A$ inherits a sparse structure from $M$ because well separated blocks do \emph{not} connect through $M$.
The operator A is implemented on a coarse linear space, and the restrictor $\PP^\dagger = \langle\phi^b_k|$ and prolongator $\PP =|\phi^b_k\rangle$ .
We can Schur decompose the matrix,
\begin{eqnarray*}
M= U D L = \left[ \begin{array}{cc}M_{\bar{s}\bar{s}} & M_{\bar{s}s} \\ M_{s\bar{s}} & M_{ss} \end{array} \right]
&=&
\left[ \begin{array}{cc} 1 & M_{\bar{s} s}  M_{ss}^{-1} \\ 0 & 1 \end{array} \right]
\left[ \begin{array}{cc} S & 0 \\ 0 & M_{ss} \end{array} \right]
\left[ \begin{array}{cc} 1 & 0 \\ M_{ss}^{-1} M_{s \bar{s}} & 1 \end{array} \right].
\end{eqnarray*}
The Galerkin oblique projectors $P_L$ and $P_R$ are formed from the diagonalisation $L$ and $U$,
\beq
P_L = P_{\bar S} U^{-1} =\left( \begin{array}{cc}
 1 & -M_{\bar{S} S}  M_{SS}^{-1}\\
 0 & 0
\end{array} \right) = (1 - MQ),
\eeq
\beq
P_R = L^{-1} P_{\bar{S}}  =
\left( \begin{array}{cc}
1 & 0 \\ -M_{SS}^{-1} M_{S \bar{S}} & 0
\end{array} \right) = (1 - QM),
\eeq
and 
\beq P_L M = \left[ \begin{array}{cc} S & 0 \\ 0 &0 \end{array} \right],    \eeq
yields the Schur complement $ S = M_{\bar{s}\bar{s}} - M_{\bar{s}s} M^{-1}_{ss} M_{s\bar{s}} $.

\subsection{Chebyshev and Krylov smoothers}

We introduce a smoother, and in this work either
use a low order Chebyshev polynomial approximation to $1/x$ over a fixed range, or use a fixed iteration count Krylov solver.
We find the Chebyshev smoother numerically more efficient than Schwarz alternating procedure (SAP) based smoothers, but they do have significantly higher
communication load when more than one processing node is used. For our small volume tests this is relatively immaterial as we
have very fast intra-node communication and use only a single node. It is worth commenting that Chebyshev functions are excellent tools, diagnostics
and probes for algorithmic issues; the ability to use a spectral band pass filter and measure the power spectrum of the residual on any given outer
iteration on a test system allows precise diagnostics of the convergence of the system and where preconditioners should be improved.
The Chebyshev polynomials are,
\beq
T_n(x) = \cos\left( n \cos^{-1} x\right).
\eeq
These can be produced with a recurrence relation.
The general interval $[a,b]$ is mapped to the standard Chebyshev interval $[-1,1]$ with the transformation $x = 2(y-a)/(b-a)-1$.
For non-hermitian systems, we base the Chebyshev smoother of degree N on the normal residual system
\beq S_{\rm chebyshev([a,b],N)}=\left[\frac{1}{2} c_0 + \sum_{j=1}^N c_j T_j(\frac{ 2 ( M^\dagger M - a) }{b-a} -1 ) \right]
M^\dagger\eeq
where the Chebyshev
coefficients $c_j$ are given by the usual Chebyshev approximation sum, taking $f(y) = 1/y$, $\theta_k = \frac{\pi (k+\frac{1}{2})}{N}$, and
\beq
  c_j = \frac{2}{N} \sum_{k=0}^{N-1} f(y(\cos \theta_k)) \cos\left( j \theta_k \right).
\eeq
$\Gamma_5$ hermiticity makes this equivalent to a polynomial of degree $2N+1$ in the hermitian indefinite matrix $M\Gamma_5$, and since $1/x$
is odd this normal residual approach is a convenient way to generate an appropriate (optimal in the Chebyshev sense) smoother polynomial.
In the case of the multigrid algorithm\cite{Brower:2020xmc}, a fixed iteration count of either GCR or
BiCGSTAB is used as a smoother.

\subsection{Composite V(1,1) multigrid preconditioner}

To maintain hermiticity in the outer iteration, we presently
introduce the smoother and coarse grid correction as preconditioner in a symmetric way, with the composite
outer Krylov operating on the matrix with a multigrid cycle as a preconditioner.
The equivalence of a sequence of multigrid correction steps to a preconditioner can be seen if consider the $V_{11}$
with a pre-smoother (S), coarse correction (Q), and post-smoother (S) in sequence,
\begin{eqnarray}
x_1 &=& x_0 + S r_0 \\
x_2 &=& x_1 + Q r_1 \\
x_3 &=& x_2 + S r_2 .
\end{eqnarray}
Since we may substitute and reduce the final update in terms of $r_0 = b - M x_0$ and $x_0$,
\begin{eqnarray}
  r_1 &=& b-M x_1 = r_0 - M S r_0 \\
  r_2 &=& b-M x_2 = r_0 - M S r_0  - M Q r_0 + M Q M S r_0 .
\end{eqnarray}
The final update sequence is then,
\begin{eqnarray}
  x_3 &=& x_0 + \left[ S (1 - MQ ) + Q + (1 - QM) S  + S( M Q M - M) S  \right]r_0\\
      &=& x_0 + \left[ S P_L + Q + P_R S  + S P_L M S  \right]r_0 .
\end{eqnarray}
This $V(1,1)$ multigrid error cycle suggests the adoption of the matrix,
\beq
\left[ S P_L + Q + P_R S  + S P_L M S  \right]
\eeq
applied to the current residual as a preconditioner in an outer Krylov solver, with its implementation
being as the above sequence of error correction steps based on the current residual $r_0$ as input.

\subsection{Deflation basis setup}
\label{sec:chebyshev}

A gauge theory is by construction invariant under local redefinitions of phase. There is necessarily no well defined
way to block or locally average degrees of freedom (without gauge fixing).
The process of averaging in this context can only be defined in a gauge covariant
way, and reduces to a classic low pass polynomial filtering problem based on eigenvalues of the (squared) gauge covariant Dirac operator.
Typically this has used (approximate) inverse iteration\cite{Luscher:2007se,Brannick:2007cc,Brannick:2007ue},
latterly with multiple passes using the multigrid solver to improve
itself\cite{Frommer:2012mv,Frommer:2013fsa,Frommer:2013kla,Richtmann:2019eyj}
We have tried several approaches to define the low mode vectors used in coarsening. These included:
\begin{enumerate}
\item Inverse iteration applied to Gaussian noise,
\item Lanczos eigenvectors, and
\item Chebyshev filters applied to Gaussian noise.
\end{enumerate}
  
We found that for the same deflation efficiency with a fixed number of vectors that Chebyshev filters had the least cost.
The rapid divergence of a high order
Chebyshev outside the default interval $[-1,1]$ is used to \emph{enhance} the modes of interest. We adopt the trick from polynomial
preconditioned implicitly restarted Lanczos \cite{rudy}.

Our setup uses the squared matrix $M^\dagger M$ with upper eigenvalue $\lambda_{\mathrm{max}}$ determined by the power method.
A low pass value is selected $\lambda_{\rm lo}$, and we use a first pass filter of order $m$ applied to Gaussian noise $\eta$,
\beq
T_m(\frac{ 2 ( M^\dagger M - \lambda_{\rm lo}) }{\lambda_{\mathrm{max}-\lambda_{\rm lo}}} -1) \eta.
\eeq
A series of vectors are then produced from this pre-filtered vector
using the Chebyshev recursion relation to generate them all for the cost of the highest order polynomial used.

\begin{figure}[hbt]
\includegraphics*[width=\textwidth]{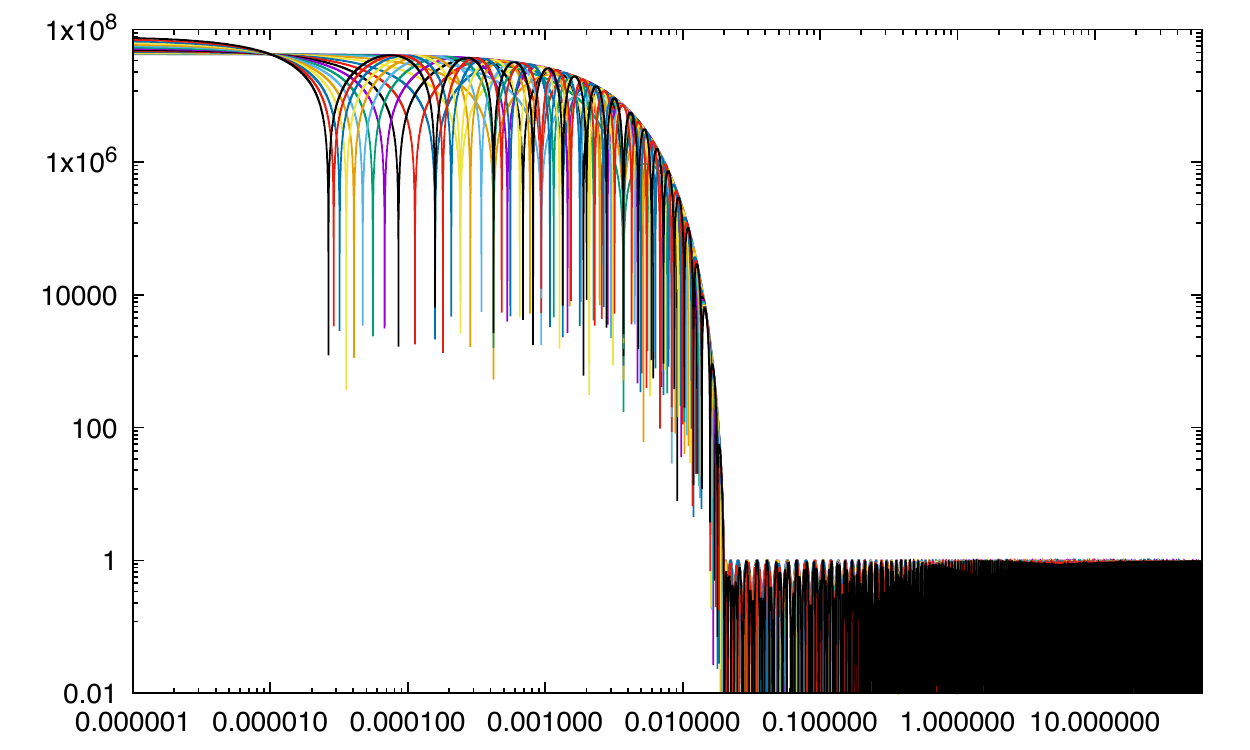}
\caption{\label{fig:chebyshev}
  Overlay of Chebyshev low-pass filter functions used to create the subspace.
  An initial O(500) low pass filter enhances the
  low mode content from Gaussian noise by $10^8$. This is
  refiltered with further, recursive Chebyshev functions of the fine operator, with a total of 1500 further matrix multiplies for 16 vectors.
  Plotted is the absolute value of the filter response for each of the 16 filtered vectors versus eigenvalue.
  The filter parameters have $\lambda_{\rm max}=60$, $m=500$, $\Delta=100$ and $\lambda_{\rm lo} = 0.02$, and are taken from the optimimum
  from a $16^3 \times 32$ domain wall fermion test case. The downward drops (obviously) correspond to the roots of the polynomials and bands within the
  target spectral range have relative signs inserted for different order polynomials, creating independent vectors with an efficient recursion.
}
\end{figure}

These later filters, based on \beq T_n(\frac{ 2 ( M^\dagger M ) }{\lambda_{\mathrm{max}}} -1),\eeq are univariate with roots at,
\beq
x_k = \frac{ 2 ( \lambda ) }{\lambda_{\mathrm{max}}} -1 = \left( \frac{\pi}{n} (k+\frac{1}{2})\right),
\eeq
giving the corresponding locations of sign changes in terms of eigenvalues of the squared fine operator. The Chebyshev
polynomial orders are chosen to build significant rearrangment of the overlap coefficients in the target deflation window.
We choose a spacing between successive polynomial orders as $ \Delta \sim O(100)$, designed to place significant degree 
of sign flipping in the important low eigenvalue region of the spectrum, between successive vectors and make
these substantially unrelated to each other. One could conjecture that the local coherence property should apply to
eigenvectors with similar eigenvalues (covariant curvature) and that this method spanning the spectrum with wholesale
negation of bands within the eigenspectrum might be efficient. Regardless of subjective interpretation, we find empirically
that the set of filtered vectors are then generated cheaply and also effective in deflation. The resulting filtered vectors,
all produced from a single initial Gaussian are,
\beq
\phi_k = 
T_{k \Delta}(\frac{ 2 ( M^\dagger M ) }{\lambda_{\mathrm{max}}} -1)
T_m(\frac{ 2 ( M^\dagger M - \lambda_{\rm lo}) }{\lambda_{\mathrm{max}}} -1)
\eta .
\eeq
An example of our use of Chebyshev filters is given in figure~\ref{fig:chebyshev}.

Alternate polynomial schemes have been tried, but without substantial gain (nor loss), for example the Chebyshev can by shifted to place the
first root at a desired place in the spectrum. Other orthogonal polynomials on the interval $[-1,1]$ are possible
choices. The Jacobi polynomials are also defined by recurrence relations, and are orthogonal under the metric,
\beq
\int\limits_{-1}^{1} dx \phi_a(x) \phi_b(x) (1-x)^\alpha (1+x)^\beta,
\eeq
where $\alpha,\beta > -1 $ and. Gegenbauer, Chebyshev, Legendre and Zernike polynomials are special cases. Varying the
weight function to place greatest weight (a positive $\beta$) at small eigenvalues was investigated, but did not so far yield a measurably more efficient
deflation space. The Chebyshev approach ($\alpha=\beta=-\frac{1}{2}$) was equally efficient, and even if not numerically different in cost, is
at least more widely known. In all cases relatively polynomial high orders are required to generate sufficient slew
to be useful in refiltering the low mode region.

Having obtained a basis that captures the near null space of the operator, the vectors are projected into left handed and right handed
chiralities, with $1\pm\gamma_5$ in the case of a four dimensional $D_W$ coarsening, and $1\pm\Gamma_5$ in the case
of five dimensional $D_{DW}$ coarsening.
This $\Gamma_5$ compatible approach was important to eliminate near zero eigenvalues in the coarsened operator with HDCR\cite{Kate}.
Faithful multiplication by $\gamma_5$ ($\Gamma_5$) is possible in the coarse space because it has a known sign
relation, the appropriate $\Gamma_5$ operator commutes with the prolongation and restriction operators, and $\Gamma_5$ hermiticity is inherited in the coarse space.

\subsection{Coarse space deflation}

On our small volume tests, we found it efficient to deflate the coarse space using eigenvector deflation. Since the coarse operator is inverted repeatedly,
the Lanczos setup overhead can be amortised and the number of required eigenvectors is limited. Since the density of eigenvectors is physical,
and the coarsening is designed to preserve low modes, we found that in our larger, $48^3$ volume, the number of eigenvectors required to deflate the coarse operator
remained large and cost is of order (Fine Volume) $\times$ (Coarse Volume). Where Lanczos was used, we make use of Chebyshev polynomial preconditioned Lanczos\cite{rudy}.
On larger volumes additional multigrid levels and also red black preconditioning were more helpful.

\section{Algorithms}
\label{sec:algorithms}

We investigated a number of different, broadly related algorithms,
based on previous work by the authors\cite{Yamaguchi:2016kop}
and the recent work reference~\cite{Brower:2020xmc}. 

{\bf Algorithms A and B} are specified in Table~\ref{tab:algorithmsAB} and are variations of HDCR, with
different coarse blocking factors ($4,2,4,2,16$ and $2^4,16$) and numbers of deflation vectors (40 and 32).
The coarsening is based on the hermitian domain wall operator $H_{DW}$.
The comparison is intended to probe the trade off between block size and number of basis vectors.
The coarse grid in Algorithm B contains 3.2 times more degrees of freedom
than Algorithm A, while the matrix multiply cost is 2.56 times more expensive.
However, the set up costs are higher for Algorithm A, with 40 vectors, and
in HMC where we aim to amortise the cost over a single solve, it is important to minimise setup cost. Numerical experiments
have been performed over a much, much larger space than presented in this table, but these figures are typical of the optimal region.
The coarse level solve was performed with Conjugate Gradients\cite{CGNR} on
the normal residual, and was deflated with Lanczos derived eigenvectors.
The setup was based on our Chebyshev polynomial filtering scheme using the
five dimensional squared $D_{DW}^\dagger D_{DW}$ operator.

\begin{table}[hbt]
\begin{tabular}{ccc}
  Algorithm & A & B \\
  \hline
  Fine   Grid & $16^3\times 32\times 16$& $16^3\times 32\times 16$ \\
  Block       & $4\times 2\times 4\times 2\times 16$& $2^4\times 16$ \\
  Coarse Grid & $4\times8\times4\times 16\times 1$& $8^3\times 16\times 1$\\
  \hline
  Outer Krylov  & ${\rm pGCR}(H_{DW})$ & ${\rm pGCR}(H_{DW})$ \\
  Basis vectors & 40 & 32 \\
  Smoother   & $S_{\rm chebyshev}([0.5,60],12)$ & $S_{\rm chebyshev}([0.5,60],12)$ \\
  \hline
  Coarsening    & $H_{DW}$ & $H_{DW}$  \\
  Coarse Solver & Deflated CGNR & Deflated CGNR \\
  Coarse Tolerance     & 0.02/0.04 & 0.02/0.04 \\
  Coarse Eigenvectors     & 48/64  & 48/128 \\
  \hline
  Subspace $\lambda_{\rm max}$ & 60.0 & 60.0 \\
  Subspace $\lambda_{\rm lo}$ & 0.05 & 0.05 \\
  Subspace $m$ & 500 & 500 \\
  Subspace $\Delta$ & 100 & 100 \\
  \hline
\end{tabular}
\caption{ \label{tab:algorithmsAB} 
  Algorithms A and B correspond to HDCR\cite{Yamaguchi:2016kop}. The comparison is intended to probe the trade off
  between block size and number of basis vectors. }
\end{table}

Table~\ref{tab:algorithmsCDE} specifies several algorithms based on four dimensional  coarsening using coarse
representations of the $D_W$ operator. In fact with appropriate
type templating it is possible in our C++ implementation to have a common implementation between the 5D portions
of the Mobius operator between the coarse and fine spaces. One must simply ensure that the fundamental chiral
projection operations are implemented on both spaces, and provide a virtual $D_W$ method.
The setup for these was based on a Chebyshev polynomial of $D_W^\dagger D_W$ with the mass coarsely tuned to be
critical by maximising the iteration count of conjugated gradients leading to ($m=-0.95$), with the first order $m=250$
Chebyshev low pass filter with $\lambda_{\rm lo}=4.0$.
The implementation was two level only and had no deflation of the coarse operator. Since the coarse space retains a
fifth dimension, it is $L_s$ fold more expensive than the coarse space for HDCR.
A new Dirac operator was implemented to produce a coarse Mobius fermion from a coarse representation of $D_W$.
The $1\pm\gamma_5$ projection of our
vectors $\phi_k$ was crucial, since $\gamma_5$ is simply a sign applied to positive and negative
chiralities, and enables the construction of a coarse space domain wall or Mobius representation from the coarsened Wilson operator.

\begin{table}[hbt]
  \begin{tabular}{ p{0.14\textwidth}ccc}
  Algorithm & C & D &E \\
  \hline
  Fine  \newline Grid & $16\times 32\times 16$& $16^3\times 32\times 16$& $16^3\times 32\times 16$ \\
  Block & $2^4\times 1$ & $2^4\times 1$& $2^4\times 1$ \\
  Coarse \newline Grid & $8^3\times 16\times 16$& $8^3\times 16\times 16$& $8^3\times 16\times 16$\\
  \hline
  Fine \newline Krylov  & ${\rm pGCR}(M_{PV}^\dagger M)$ & ${\rm pGCR}(M_{PV}^\dagger M)$ & ${\rm pGCR}(M)$ \\
  Smoother   & $S_{\rm BiCGSTAB}(24)$ &   $S_{\rm GCR}(14)$ & $S_{\rm chebyshev}([0.5,60],14)$ \\
  \hline
  Coarsening    & $D_{W}$ & $D_{W}$ & $D_{W}$  \\
  Coarse  \newline Solver & ${\rm BiCGSTAB}(M_{PV}^\dagger M)$ &  ${\rm GCR}(M_{PV}^\dagger M)$ & ${\rm CGNR}(M^\dagger M)$ \\
  Coarse \newline  Tolerance     & 0.02 & 0.02 & 0.02/0.1 \\
  Coarse  \newline Eigenvectors     & 0 & 0 & 0/64 \\
  \hline
  Subspace  \newline basis   & 24 & 24  & 24 \\
  Subspace $\lambda_{\rm max}$ & 60.0 & 60.0  & 60.0 \\
  Subspace $\lambda_{\rm lo}$ & 4.0 & 4.0 & 4.0 \\
  Subspace $m$ & 600 & 600 & 600 \\
  Subspace $\Delta$ & 250 & 250 & 250 \\
  \hline
\end{tabular}
\caption{ \label{tab:algorithmsCDE} 
  Algorithm C corresponds to reference \cite{Brower:2020xmc}, but uses a GCR outer solver.
  The smoother was a fixed number of iterations of BiCGSTAB, and we found that O(24) were required
  to maintain convergence of the solution. Algorithm D replaces the BiCGSTAB smoother with a GCR smoother,
  and we found convergence remained possible with reduced smoother cost.
  Algorithm E uses the 4D coarsening scheme of reference \cite{Brower:2020xmc}, but combined with the 
  multigrid approach of HDCR\cite{Yamaguchi:2016kop}. The deflation subspaces are identical in both these
  cases. The dimension of the coarse space is proportional to $L_s$ and so substantially greater than the HDCR case.
  Algorithm E was tried both with and without eigenvector deflation in the coarse space.
}
\end{table}

{\bf Algorithm C} implements the $D_W$ based coarsening and GCR outer solver and BiCGSTAB coarse space
solver and smoother.  It is based on the Pauli Villars preconditioning scheme, with coarse operator,
$$\PP^\dagger M_{PV}^\dagger \PP \PP^\dagger M_l \PP,$$ and is similar to the algorithm
introduced in reference~\cite{Brower:2020xmc}.
A 5D coarse DWF dirac operator is created from the 4D coarse space representation of $D_W$.

{\bf Algorithm D} is almost the same as Algorithm C, but
substitutes a fixed number of GCR iterations as the smoother.

Reference~\cite{Brower:2020xmc} called these algorithms domain wall multigrid.
However, since there were three prior multi-level algorithms for domain wall fermions \cite{Cohen:2012sh,Boyle:2014rwa,Yamaguchi:2016kop}, and
two of which \cite{Boyle:2014rwa,Yamaguchi:2016kop} obtained real time to solution acceleration in four dimensional
QCD, we do not feel this naming is sufficiently specific. For the purposes of this paper we will refer to Algorithms C and D
as MG-PV, since the unique attributes are the use of a Pauli Villars left preconditioner and the use of a four dimensional $D_W$ based
coarsening.

{\bf Algorithm E} is new scheme, using the $D_W$ based coarsening of reference~\cite{Brower:2020xmc},
with multigrid preconditioned GCR on $D_{DW}$ operator performed at the fine level, and
with Conjugate Gradients solution of the $M^\dagger_l M_l$ system in the coarse
space. This represents $D_W$ coarsening of reference~\cite{Brower:2020xmc},
but combines it with a squared coarse operator
$\PP^\dagger M^\dagger_l \PP \PP^\dagger M_l \PP$ instead of $\PP^\dagger M_{PV}^\dagger\PP \PP^\dagger M_l\PP$,
thereby being similar in this respect to HDCR \cite{Yamaguchi:2016kop} but with the 4D coarsening of the Wilson operator from\cite{Brower:2020xmc}.
Since both the coarse and fine matrices are hermitian positive definite, we can use Lanczos derived eigenvector deflation of the second level,
and this accelerates the coarse space convergence. We make a direct comparison between the solution with, and without,
eigenvector deflation of the coarse space. We will refer to this algorithm as MG-$M^\dag M$.

{\bf Algorithm F}, Table~\ref{tab:AlgorithmF},
is a hybrid three level scheme combining the fast (four dimensional) setup in the fine
space with coarsening using a representation of $D_W$ but stepping through a five dimensional setup in the
coarse space to a third, coarse-coarse, level using a represention of $H_{DW}$ on the coarse space. A fixed iteration count W-cycle is used with
no convergence precision constraint on the intermediate level solver, so that two coarse-coarse corrections
are used for every coarse grid step.
We will refer to this algorithm as Hybrid-$M^\dag M$.

\begin{table}[hbt]
\begin{tabular}{cc}
  Algorithm & F \\
  \hline
  Fine   Grid & $16^3\times 32\times 16$ \\
  \hline
  Fine Krylov  & ${\rm pGCR}(M)$ \\
  Smoother     & $S_{chebyshev}([0.5,60],12)$ \\
  \hline
  Coarsening    & $D_{W}$  \\
  Coarse Grid & $8^3\times 16\times 16$\\
  Coarse Solver        & pGCR(M) \\
  Coarse Tolerance     & 0.01 \\
  \hline
  Subspace basis   & 24 \\
  Subspace $\lambda_{\rm max}$ & 60.0  \\
  Subspace $\lambda_{\rm lo}$ & 4.0  \\
  Subspace $m$ & 600  \\
  Subspace $\Delta$ & 250 \\
  \hline
  Coarsening    & $H_{DW}$  \\
  Coarse Grid & $8^3\times 16\times 1$\\
  Coarse Solver & $CGNR(M^\dagger M)$ \\
  Coarse Tolerance     & 0.02 \\
  Coarse Eigenvectors     & 128 \\
  \hline
  Subspace basis   & 32 \\
  Subspace $\lambda_{\rm max}$ & 60.0  \\
  Subspace $\lambda_{\rm lo}$ & 4.0  \\
  Subspace $m$ & 600  \\
  Subspace $\Delta$ & 250 \\
  \hline
\end{tabular}
\caption{ \label{tab:AlgorithmF} 
  {Algorithm F} is a hybrid three level scheme combining one level of coarsening using a
  representation of $D_W$  with a second level of coarsening using $H_{DW}$ on the coarse space.
  The algorithm is then deflated using Lanczos derived eigenvectors in the coarse-coarse space and
  CGNR.
}
\end{table}

{\bf Algorithm G}, Table~\ref{tab:AlgorithmG}, is a reimplementation of HDCG\cite{Boyle:2014rwa} using the next-next-next-nearest-neighbour
  squared red-black Schur complement operator $M_{pc}^\dagger M_{pc}$. The algorithm is sub-optimally implemented
  in our analysis since the coarse operator is constructed via application of the fine operator
  and projection back to the coarse space. This is because the code to implement the 81 point stencil is
  tedious to reimplement, but it is useful to include this for in principle algorithm performance comparisons, even
  if the timings are poor. Where we later quote fine matrix multiple counts, we do not include those multiplies
  made in applying the coarse operator, since this is artificially introduced as an artefact of the man-power
  efficient, but computer time inefficient implementation. 
  
\begin{table}[hbt]
\begin{tabular}{cc}
  \hline
  Fine   Grid & $16^3\times 32\times 16$ \\
  \hline
  Fine Krylov  & ${\rm pCG}(M_{ee} - M_{eo} M_{oo}^{-1} M_{oe})^\dagger(M_{ee} - M_{eo} M_{oo}^{-1} M_{oe})$ \\
  Smoother     & $S_{chebyshev}([0.5,60],12)$ \\
  \hline
  Coarsening    & $(M_{ee} - M_{eo} M_{oo}^{-1} M_{oe})^\dagger(M_{ee} - M_{eo} M_{oo}^{-1} M_{oe})$  \\
  Coarse Grid & $8^3\times 16\times 1$\\
  Coarse Solver        & CG \\
  Coarse Tolerance     & 0.05 \\
  Coarse Eigenvectors     &  32 \\
  \hline
  Subspace basis   & 32 \\
  Subspace $\lambda_{\rm max}$ & 30.0  \\
  Subspace $\lambda_{\rm lo}$ & 0.02  \\
  Subspace $m$ & 500  \\
  Subspace $\Delta$ & 100 \\
  \hline
\end{tabular}
\caption{  \label{tab:AlgorithmG} 
  Algorithm G is an arithmetically correct, but inefficient, reimplementation of HDCG\cite{Boyle:2014rwa}.
}
\end{table}


\section{Results}

\label{sec:numerics}

A head-to-head comparison between algorithms was performed
on a single $16^3\times 32$ domain wall fermion configuration number 4000 from our $m_{ud}=0.01$ 2+1 flavour Iwasaki gauge
ensemble at $\beta=2.13$\cite{Allton:2007hx}. The valence quark mass was set to the non-unitary value of $m_q=0.001$ to increase the condition number of the linear system.
All code has been implemented in the Grid library\cite{Boyle:2016lbp}.
The Summit system at Oak Ridge National Laboratory has been
used for the code development, testing and timings. All code was run in exclusively double precision
for simplicity and in order to separate algorithmic convergence rate from any numerical precision optimisations.
Single precision and mixed precision code was also run, and shared the usual multigrid property of yielding a more precise
true residual than a single precision red black preconditioned Krylov solver.

\subsection{Conventional Krylov Solvers}

\begin{figure}[hbt]
\includegraphics*[width=\textwidth]{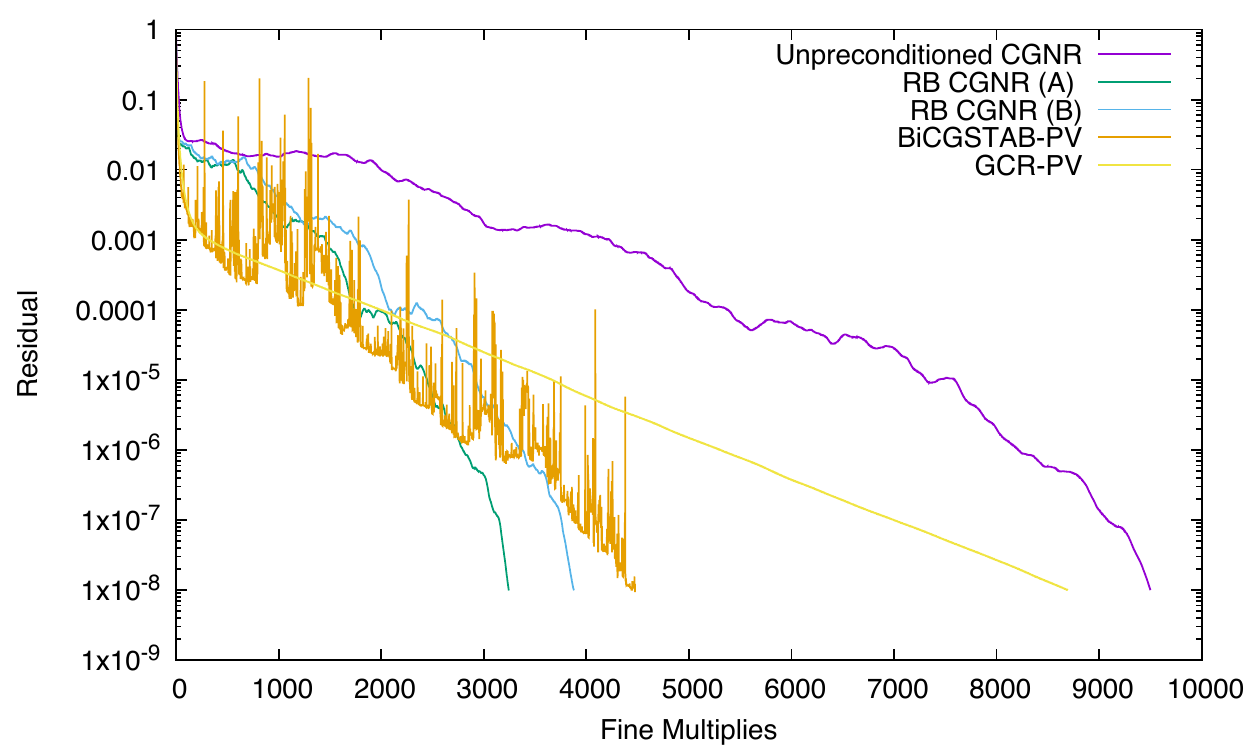}
\caption{\label{fig:Fine}
  We plot the convergence history of the five conventional Krylov solver algorithms, from Table~\ref{tab:krylovs}.
}
\end{figure}

The baseline against which to demonstrate benefit
for our multigrid solvers is the \emph{best} standard Krylov solver
algorithm. For our test system show the number of fine matrix multiplies required for each of
CGNR, red-black preconditioned CGNR, and both BiCGSTAB and GCR on the fine $M_{PV}^\dagger M_l$ operator.
This latter is included to show the algorithmic impact of the spectral transformation using Pauli Villars preconditioning,
introduced in\cite{Brower:2020xmc}. In Table~\ref{tab:krylovs} we see that at this input quark mass the
Pauli Villars preconditioner produces some benefit relative to the unpreconditioned system, reducing the iteration count
from 9541 to 4140, and the number of full grid fine matrix multiplies from 19082 to 8280 compared to Conjugate Gradients.
This is a clear benefit in condition number arising from not squaring the smallest eigenvalue, as anticipated in the
introductory discussion of the spectrum.
Figure~\ref{fig:Fine} compares the convergence history of these algorithms.
Although GCR has an inefficient convergence history, the early convergence was sufficiently reasonable that
we also adopt a fixed iteration count GCR based smoother option, in addition to the BiCGSTAB smoother proposed in \cite{Brower:2020xmc}.

At sufficiently light quark masses, this effect should win over red black preconditioning, but the cross over is not
obtained in practice on our test systems, since standard red black preconditioning gives a similar benefit at this
quark mass. Each iteration of the even-odd solver has four
applications of the even odd hopping term, costing half the floating point operations of a full grid matrix multiply.
Each iteration is therefore equivalent to two both checkerboard matrix multiplies, and we count in the units of both
checkerboards to keep the number of matrix multiplies directly comparable to 
the non-red-black solvers. In other words, $(M_{ee} - M_{eo} M_{oo}^{-1} M_{oe})$ is counted as a single both-checkerboard
matrix multiply.
The corresponding cost for red-black solves are therefore 7760 and 6448 matrix multiplies and 3880 and 3224 iterations depending on the precise
details of the Schur decomposition scheme used in the preconditioner.

\begin{table}[hbt]
\begin{tabular}{ccccc}
  \hline
  Algorithm & Operator & Iterations& Full Matmuls & Time (s)\\
  \hline
CGNR      &  $M^\dagger M$  & 9541 &  19082& 183s \\
BiCGSTAB  &  $M_{PV}^\dagger M$ & 4140 &  8280   & 79s\\
prec-CGNR   &  $(M_{ee} - M_{eo} M_{oo}^{-1} M_{oe})^\dagger(M_{ee} - M_{eo} M_{oo}^{-1} M_{oe})$   & 3224  & 6448  & 62s\\
prec-CGNR   &  $(1 - M_{ee}^{-1}M_{eo} M_{oo}^{-1} M_{oe})^\dagger(1 - M_{ee}^{-1}M_{eo} M_{oo}^{-1} M_{oe})$   & 3880  & 7760  & 77s\\
GCR(32,32)  &  $M_{PV}^\dagger M$ & 8693 &  17386   & 474s\\
\hline
\end{tabular}
\caption{ \label{tab:krylovs} We display the wall clock timing, iteration count and (full grid)
  matrix multiply count for each
  of unpreconditioned Conjugate Gradients, BiCGSTAB on the Pauli Villars preconditioned operator,
  two forms of red-black preconditioned
  Conjugate Gradients. In the two red black preconditioned cases the number of half grid matrix multiplies
  is double the full grid count given in this table, and the cost of each is of course halved, making the
  all the counts in this table an equivalent cost comparison.
}
\end{table}

\subsection{Numerical cost of multigrid algorithms}

For each algorithm, the parameters were carefully tuned and are believed optimal
within the space covered by the algorithm.
Multigrid parameter optimisation
is tedious, and once sensible ranges of each parameter were identified, these were looped over in a brute force program
and the best case selected.
These results are central to this paper, and require some careful discussion. 
Table~\ref{tab:AlgorithmNames} gives a key to the mapping of Algorithms A-G for which results are presented
to meaningful classification names. These names are explained below.

Table~\ref{tab:matrix_multsAB} displays the inner, outer iteration counts and the total number of fine
matrix multiplies for the HDCR Algorithms A, B.
Table~\ref{tab:matrix_multsCDE} displays the inner, outer iteration counts and the total number of fine
matrix multiplies for the MG-PV Algorithms C, D and the 4D coarsened MG-$M^\dagger M$ Algorithm E.
Table~\ref{tab:matrix_multsFG} displays the inner, outer iteration counts and the total number of fine
matrix multiplies for the Hybrid-$M^\dagger M$ Algorithm F and the HDCG Algorithm G.
\begin{table}[hbt]
\begin{tabular}{cc}
  \hline
  Name & Algorithms \\
  \hline
  HDCR   & A,B\\ 
  MG-PV  & C,D\\
  MG-$M^\dagger M$ &E \\
  Hybrid-$M^\dagger M$ &F \\
  HDCG &G\\
  \hline
\end{tabular}
\caption{  \label{tab:AlgorithmNames} 
  Names used for different classes of algorithm in this paper.
}
\end{table}

\begin{table}[hbt]
  \begin{tabular}{
      p{0.12\textwidth}
      p{0.12\textwidth}
      p{0.12\textwidth}
      p{0.12\textwidth}
      p{0.12\textwidth}
      p{0.12\textwidth}
      p{0.12\textwidth}
    }
\hline
 &   Coarse\newline Iterations & Outer\newline Iterations& Fine\newline Matmuls & Time & Coarse\newline residual & Coarse\newline eigenvectors\\
\hline
RB-CGNR   &     -      &  3224     &  6448   &   62s    & - & - \\ 
\hline
A      &     83      &  25        &  1200    &   20.7s  & 0.02& 48  \\ 
A      &     66      &  24        &  1152    &   19.6s  & 0.02& 64  \\ 
A      &     54      &  25        &  1200    &   19.8s  & 0.04& 64  \\ 
A      &     34      &  28        &  1344    &   21.8s  & 0.1 & 64  \\ 
A      &     47      &  23        &  1104    &   18.6s  & 0.02& 128 \\ 
A      &     40      &  24        &  1152    &   19.1s  & 0.04& 128 \\ 
A      &     27      &  24        &  1152    &   18.8s  & 0.1& 128  \\ 
\hline
B      &     94      &  22        &  1056    &   20.5s  & 0.02& 48  \\ 
B      &     58      &  21        &  1008    &   18.8s  & 0.02& 128 \\ 
B      &     44      &  21        &  1008    &   17.7s  & 0.04& 128 \\ 
B      &     66      &  24        &  1152    &   19.8s  & 0.1 & 128 \\ 
\hline
\end{tabular}
\caption{ \label{tab:matrix_multsAB} We display the wall clock timing and matrix multiply count for
  HDCR Algorithms A, B. The coarse iterations are the number of iterations per coarse solve, and this is
  performed once every outer iteration in each algorithm. The total number
  of coarse matrix applications are therefore the product of these numbers. The number of fine matrix
  multiplies are directly comparable in all cases.
  The lattice volume is $16^3\times 32$ and the (non-unitary) light quark
  mass is $m_l=0.001$. The coarsening factor is $2^4\times 16$, and 16 left-handed and 16 right-handed basis vectors
  are used.
}
\end{table}

\begin{table}[hbt]
  \begin{tabular}{
      p{0.12\textwidth}
      p{0.12\textwidth}
      p{0.12\textwidth}
      p{0.12\textwidth}
      p{0.12\textwidth}
      p{0.12\textwidth}
      p{0.12\textwidth}
    }
\hline
Algorithm &  Coarse\newline Iterations & Outer\newline Iterations& Fine\newline Matmuls & Time & Coarse\newline residual & Coarse\newline eigenvectors\\
\hline
RB-CGNR   &     -      &  3224     &  6448   &   62s    & - & - \\ 
\hline
C      &     254     &  40        &  3840    &   1387s  & 0.02 & -  \\ 
C      &     208     &  40        &  3840    &   1089s  & 0.04 & -  \\ 
C      &     100     &  -         &  -       &   $\infty$ & 0.1 & -   \\ 
\hline
D      &     307     &  49        &  2744    &   1904s  & 0.02 & -   \\ 
D      &     184     &  51        &  2856    &   1554s  & 0.04 & -   \\ 
D      &     161     &  65        &  3640    &   1584s  & 0.1  & -   \\ 
\hline
E      &     554     &  35        &  1960    &   1051s  & 0.02 & -    \\ 
E      &     517     &  49        &  2744    &   1253s  & 0.04 & -    \\ 
E      &     365     &  72        &  4032    &   1620s  & 0.1 & -    \\ 
E      &     127     &  34        &  1904    &   298s   & 0.02& 64     \\ 
E      &      98     &  33        &  1932    &   235s   & 0.04& 64     \\ 
E      &     58      &  27        &  1512    &   143s   & 0.1 & 64   \\ 
\hline
\end{tabular}
\caption{ \label{tab:matrix_multsCDE} We display the wall clock timing and matrix multiply count for
  Algorithms C, D and E. The coarse iterations are the number of iterations per coarse solve, and this is
  performed once every outer iteration in each algorithm. The total number
  of coarse matrix applications are therefore the product of these numbers. The number of fine matrix
  multiplies are directly comparable in all cases.
  The lattice volume is $16^3\times 32$ and the (non-unitary) light quark
  mass is $m_l=0.001$. The coarsening factor is $2^4\times 1$, and 12 left handed and 12 right handed basis vectors
  are used.
}
\end{table}

\begin{table}[hbt]
  \begin{tabular}{
      p{0.12\textwidth}
      p{0.12\textwidth}
      p{0.12\textwidth}
      p{0.12\textwidth}
      p{0.12\textwidth}
      p{0.12\textwidth}
      p{0.12\textwidth}
    }
\hline
Algorithm &  Coarse\newline Iterations & Outer\newline Iterations& Fine\newline Matmuls & Time & Coarse\newline residual & Coarse\newline eigenvectors\\
\hline
RB-CGNR   &     -      &  3224     &  6448   &   62s    & - & - \\ 
\hline
\hline
F      &     56      &  29        &  1392    &    54s   & 0.02 & 128 \\ 
\hline
G      &     53      &  26        &   728    &   127s & 0.05 & 32    \\ 
\hline
\end{tabular}
\caption{ \label{tab:matrix_multsFG} We display the wall clock timing and matrix multiply count for
  Algorithms F and G. The coarse iterations are the number of iterations per coarse solve, and this is
  performed once every outer iteration in each algorithm. Algorithm F uses a fixed W-cycle
  and performs two solves in the doubly coarsened space for every outer iteration. The number of fine matrix
  multiplies are directly comparable in all cases.
  The lattice volume is $16^3\times 32$ and the (non-unitary) light quark
  mass is $m_l=0.001$.  The coarsening factors is $2^4\times 1$ and $1^4\times 16$ for three level Algorithm F,
  and $2^4\times 16$ for Algorithm G.
  16 left-handed and 16 right-handed basis vectors are used for Algorithm F and 32 basis vectors are used for Algorithm G.
}
\end{table}

\begin{table}[hbt]
  \begin{tabular}{
      p{0.12\textwidth}
      p{0.12\textwidth}
      p{0.12\textwidth}
      p{0.12\textwidth}
      p{0.12\textwidth}
      p{0.12\textwidth}
      p{0.12\textwidth}
    }
\hline
Algorithm &  Coarse\newline Iterations & Outer\newline Iterations& Fine\newline Matmuls & Time & Coarse\newline residual & Coarse\newline eigenvectors\\
\hline
red-black\newline CGNR &  - &  3224     &  6448   &   62s    & - & - \\ 
\hline
C      &    90 & 107 & 12840 & 790s & 0.02 & - \\ 
D      &    97 & 98  & 6272 & 805  & 0.02 & - \\ 
E      &    63 &  68 & 3264  & 164s & 0.1  & 64 \\ 
%
\hline
\end{tabular}
\caption{ \label{tab:matrix_mults_4444} We display the wall clock timing and matrix multiply count for
  Algorithms C, D and E with a block size of $4^4$, rather than the $2^4$ previously, and an increase in the
  number of basis vectors from 24 to 48. In the case of algorithms $C$ based on the non-hermitian
  $M_{PV}^\dagger M$ system, the order of smoother had to be increased from $24$ to $30$ to retain convergence.
  The lattice volume is $16^3\times 32$ and the (non-unitary) light quark
  mass $m_l=0.001$. The coarsening factor is $4^4\times 1$, 24 left-handed and 24
  right-handed basis vectors are used.
}
\end{table}

{\bf Algorithms A and B} have been fully optimised in terms of code execution, and achieve a three fold CPU time speed up compared
to red-black preconditioned CGNR and ten fold speed up compared to unpreconditioned CGNR.
Further this is done with the optimisation being chosen to minimise the sum of set up and solve times,
rather than the solve time in isolation. The aim was to produce a viable algorithm for application in Hybrid Monte Carlo
evolution.

The total matrix multiply count with Algorithms A and B (HDCR)
is 45\% more than achieved with Algorithm G (HDCG), while the outer iterations, indicative of deflation
efficacy, are similar. The difference largely arises from the use of the ADEF2 two level CG algorithm\cite{Boyle:2014rwa} in HDCG,
where the smoother is applied once per iteration. We empirically discover that a similar order smoother can be
maintained reducing the overall smoothing effort.
The smaller spectral range of the preconditioned operator likely contributes to this
reduction in required smoothing effort. Combining red-black preconditioning in a smoother with the
our $H_{DW}$ solver was not effective, likely because the eigenvectors of $H_{DW}$ and the red-black
operator are not aligned, and having carefully eliminated the low mode error with a coarse grid correction, immediately reintroducing them
with a smoother is detrimental.

Algorithms A and B are thus reasonably competitive to HDCG,
while being more viable for application in the Hybrid Monte Carlo algorithm.
Our HDCG re-implementation is (quite deliberately) sub-optimal in terms of wall clock, with the
non-local coarse operator implemented via promotion to the fine space, fine matrix multiplications, and restriction
to coarse space. Despite this inefficient implementation, the HDCG time to solution is moderately competitive, because it has
the modest fine matrix multiply counts and a deflated coarse space.
However, Algorithm G was also somewhat artificial as a blocking factor of only $2^4$ is not possible with the
efficient implementation.
This unrepresentative time is included for complete information. What is
important is to show that the HDCR and HDCG algorithms are broadly similar in
fine matrix multiply counts, showing that the multigrid preconditioners have similar efficiency.

{\bf Algorithms C, D (MG-PV) and E (MG-$M^\dagger M$)} use coarsening in four dimensions
only. The coarse spaces are five dimensional,  and comparing them raises several interesting points of note. 
Firstly for Algorithm C, the smoother was a fixed number of iterations of BiCGSTAB,
and we found that a larger number of smoother matrix multiplies
O(24) were required to maintain convergence of the solution.
GCR was found to be a better smoother in Algorithm D, both maintaining convergence with fewer smoother iterations,
even lower than 14, and obtaining faster convergence rate at the same number of smoother iterations compared to Algorithm C.
Algorithms C, D, E and F successfully apply $D_W$ based coarsening in 4D QCD, demonstrated previously in the 2D Schwinger model\cite{Brower:2020xmc},
is significant because this is based on a four dimensional set up with the Wilson operator
rather than the five dimensional matrix being inverted. The setup on a lower dimensional space is cheap, and 
the idea works in principle. However we have not yet made it give a compelling gain.

Algorithms C and D (MG-PV) have poorer deflation efficiency than both HDCR and HDCG algorithms in terms of fine matrix multiplies.
However this still represents an over two-fold gain compared to the
original red-black CGNR Krylov solver and six-fold compared to the unpreconditioned
CGNR algorithms, when measured by fine matrix multiplies. This is not dissimilar to the ratio seen in the original
2D Schwinger model results \cite{Brower:2020xmc}. We use fewer basis vectors here, and smaller blocking factors of size $2^4$.
However, the coarse space retains a fifth dimension of size $L_s$, and the
greater size of this space compared to Algorithms A and B, and this
makes the cost of the coarse space significant and leads to a longer run time for the algorithm.
The potential gain is lost.

The coarse solves in cases C and D are not deflated.
Since Algorithms C and D are based on a non-hermitian matrix one would have to
solve the pseudospectrum and use a Singular Value Decomposition
to handle the non-normal case, but we have not implemented this algorithm in our code.
In the absence of such deflation one should assume that the coarse space could
be reduced in cost by a factor of order ten, based on experience of the other algorithms.
However, since Algorithm E was faster when undeflated than Algorithms C and D, and a deflated variant Algorithm E was uncompetitive,
we do not see any realistic opportunity for Algorithms C and D to become competitive under deflation.

We focused on attempts to reduce the excessive cost of the coarse space.
Possibilities studied were greater blocking factors, deflation of the coarse space, recursive multigrid, reduction of the fifth dimension
in the coarse space and complete removal of the fifth dimension in the coarse space.

{\bf Blocking:}
We repeated the tuning for Algorithms C, D  and E with a larger blocking
factor of $4^4$, with result in Table~\ref{tab:matrix_mults_4444}. The number of basis vectors was simultaneously
increased from 24 to 48, so that the dimension of the coarse space vectors was eight times smaller, and the cost of the
coarse matrix multiply was only half the cost of the $2^4$ coarsening.
  We see that the outer iteration count grows, and in the case of algorithms $C$ based on the non-hermitian
  $M_{PV}^\dagger M$ system, the order of smoother had to be increased from $24$ to $30$ to retain convergence.
  Although this reduces the total cost of the algorithms in wall clock time due to the cheaper coarse space,
  the loss of deflation effectiveness
  is shown by the rise in the number of fine matrix multiplies.
  Since the number of fine matrix
  multiplies ceases to beat the red-black conjugate gradient algorithm, one sees that even if the coarse space
  were reduced to near zero cost it would still not be possible to obtain an overall speed up with this level of blocking.
  We conclude that it is necessary to use only a modest first level of coarsening for a viable
  algorithm when based on the $D_W$ coarsening. If these algorithms are to succeed, multiple levels (which we will use
  in Algorithm F) and/or eigenvector deflation likely will be required to reduce the coarse space cost.
  However, we see in Algorithm E that although deflation helped, it did not help sufficiently to match either the HDCR Algorithm A and B,
  or the red black preconditioned CGNR.

{\bf Deflation:}
Algorithm E, MG-$M^\dagger M$ was introduced using the squared operator with $D_W$ coarsening. It is
worth noting that Algorithm E was faster than Algorithms C and D (MG-PV) without deflation, both in wall clock
execution time and in terms of fine matrix multiplications. For Algorithm E,
a comparison is given between undeflated and deflated coarse space solution can be made,
and this gained a factor of 7.3. Since Algorithm E remains slower than red black preconditioned CGNR,
it requires a significant gain for Algorithms C or D to become competitive with HDCR (Algorithms A and B).

The comparison of deflated with undeflated solves in Algorithm E is interesting: the effect of coarse space deflation
with a residual of $r\le 0.1$ is not restricted to accelerating the coarse solve. The exact eigenvector deflation
reduces the outer iteration count from 34 to 27, in \emph{addition} to reducing the coarse space solve time. We might
conjecture that being exact in the lowest modes of the coarse space, which multigrid is carefully designed to preserve from the fine space,
has a benefit beyond that indicated by the relaxed convergence residual. It is possible that eigenvector deflation in the coarse
space enhances the degree to which a coarse grid correction is differentially correct in the low mode region: this
is precisely the region where the coarse representation of the fine matrix is designed to most effective in accelerating outer convergence.

{\bf Reduced $L_s$:}
Reference \cite{Brower:2020xmc} suggested using distinct $L_s$ on multiple grids and communicating only surface slices.
We found moderate effectiveness in dropping from $L_s=16$ to $L^\prime_s=8$ on the coarse grid, and that a better approach
was to communicate bands of depth $L_s^\prime/2$ nearest the surfaces between these grids. However the deflation efficacy
was reduced and the direction did not appear worth pursuing, unless other significant new ideas are composed with it.
In principle this approach could be combined with a change between a Mobius and a standard Shamir domain wall fermion formulation,
and may enable combining a Mobius fine action with coarse level approaches that are more constrained in the form of the action.

{\bf Removed $L_s$:}
Since deflation in the coarse space is estimated to be insufficient to make Algorithms C and D (MG-PV) competitive (by comparison with
Algorithm E, MG-$M^\dagger M$), we instead focus with Algorithm F on reducing the dimension and cost of the coarse space by other means. If we again restrict
ourselves to a Shamir type action, and use a second level of blocking by the full fifth dimension we can
introduce a W-cycle and spend most effort on the a coarsest Grid with a composite $2^4\times 16$ blocking and
in a coarse space that is only four dimensional. This is a hybrid three level scheme with one level of $D_W$ coarsening,
and a second level of $H_{DW}$ coarsening. Algorithm F does obtain a modest speed up compared to red black CGNR, but is
not as effective as Algorithms A and B (HDCR).

Algorithm F, Hybrid-$M^\dagger M$, was the best of the approaches we studied making use of four dimensional, $D_W$ based coarsening.
In principle this approach could be combined with a change between a Mobius and a standard Shamir domain wall fermion formulation,
and may enable combining a Mobius fine action with coarse level approaches that are more constrained in the form of the action.

\subsubsection{Convergence history}

\begin{figure}[hbt]
\includegraphics*[width=\textwidth]{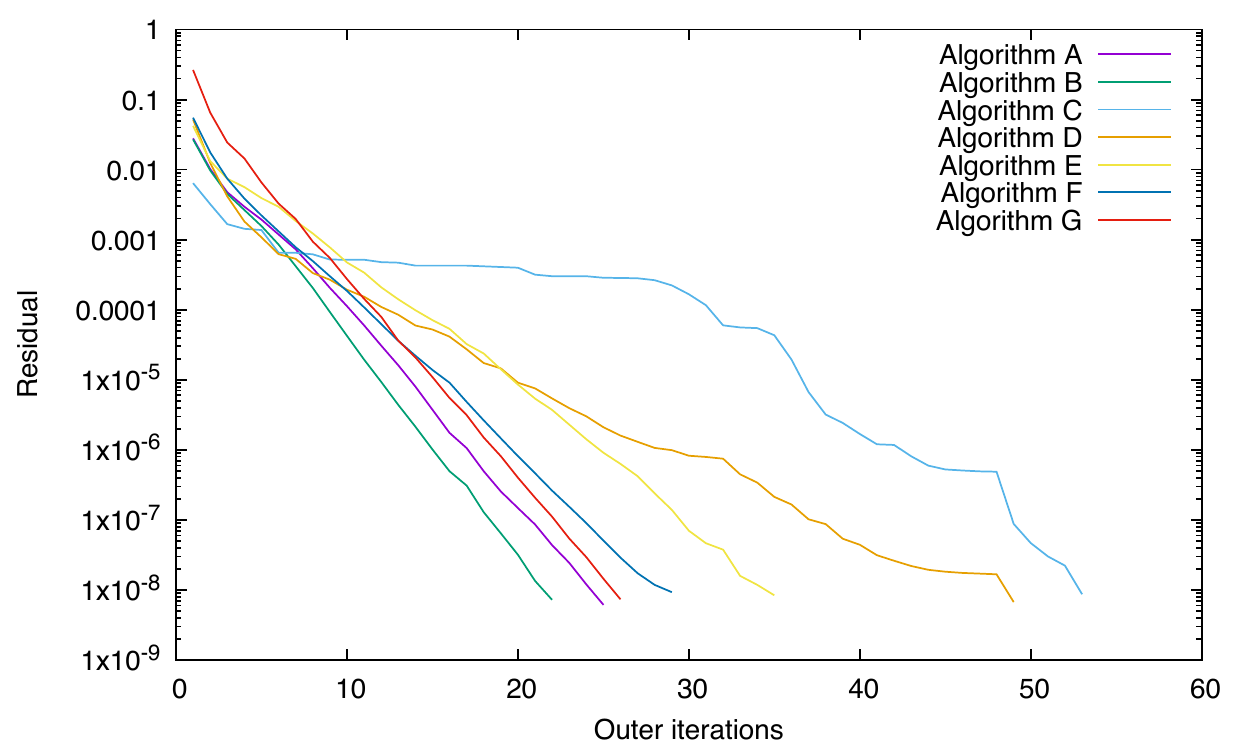}
\caption{\label{fig:Outer}
  We plot the convergence history of Algorithms A-G versus outer iterations.
  Since the smoother and coarse grid cost are not included the x-axis in anyway.
  Although convergence rate is dictated by the overall efficiency of the combined multigrid preconditioner,
  the cost of the preconditioner (which comprises both Smoothers and a coarse grid correction) must be included
  for meaningful comparison.
}
\end{figure}

\begin{figure}[hbt]
\includegraphics*[width=\textwidth]{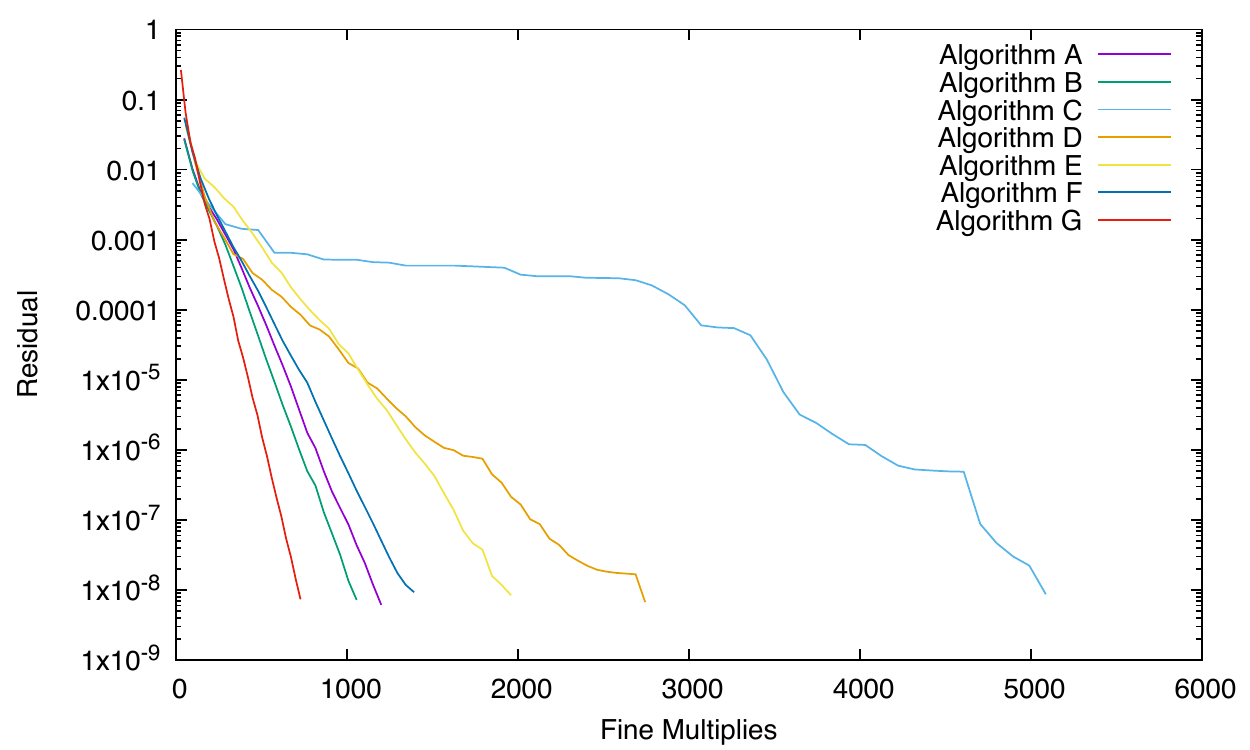}
\caption{\label{fig:Multiplies}
  We plot the convergence history of Algorithms A-G versus fine matrix multiplies.
  This includes the smoother cost, but not the cost of coarse corrections.
  Since the dimension of the coarse grid operator changes between algorithms,
  this is not yet a direct comparison. However, since we use deflated solves in the coarse
  space for Algorithms A,B, E, F and G, but not for Algorithms C and D, the irreducible cost
  given this plot will bound the cost of Algorithm C and D relative to that for the other algorithms.
  We see that Algorithms C and D take the most fine matrix multiplies, even when the coarse space is (incorrectly) assumed
  to be free. Since Algorithms C, D and E have only five dimensional coarse spaces, these algorithms would have intrinsically
  more expensive coarse spaces even if they were deflated.
}
\end{figure}

\begin{figure}[hbt]
\includegraphics*[width=\textwidth]{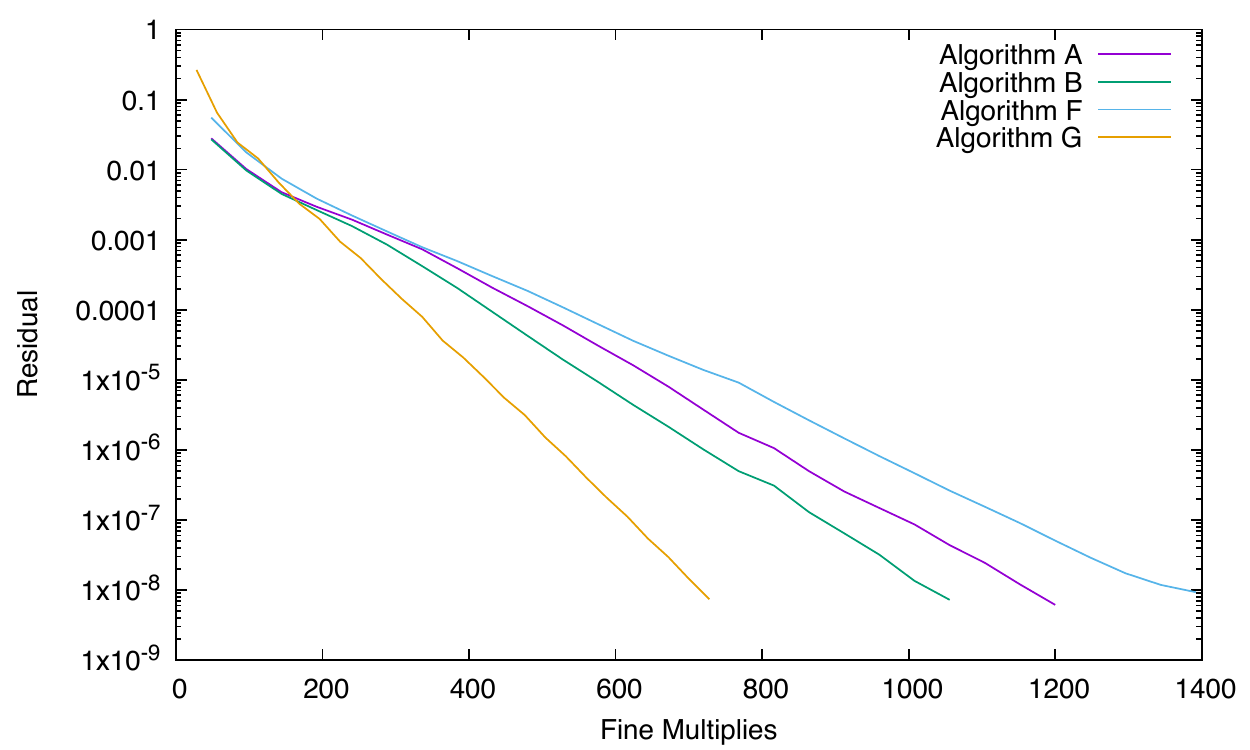}
\caption{\label{fig:MultipliesFast}
  We plot the convergence history of Algorithms A, B, F and G  versus fine matrix multiplies.
  This plots the same data as figure~\ref{fig:Multiplies} with Algorithms C, D and E removed to expand the scale for clarity.
}
\end{figure}

It is useful to present some more detail of convergence histories.
Figure~\ref{fig:Outer} displays the detailed convergence history of each of Algorithms A-G versus outer iteration number,
while Figure~\ref{fig:Multiplies} displays this same history versus fine matrix multiplies, adding the cost of multigrid smoothers.
Figure~\ref{fig:MultipliesFast} displays the same data as Figure~\ref{fig:Multiplies} but zooming and restricting the the faster
Algorithms A, B, D, E, F and G.

\subsubsection{Execution time}

When the cost of the coarse space is factored, a convergence versus time is the most useful comparison, so
we will discuss the relative execution time of the algorithms. The optimal algorithm
is computing technology dependent, and subject to variations in architectures as the
balance between the cost of the coarse grid solves and the fine grid operator will change.
While it may be attractive to try to abstract cost and develop algorithms without reference to computing
hardware, this is a fallacy. We try in this paper to focus as much as possible on machine independent statements.
However two important points are: a) the relative cost of coarse and fine space operations is dictated by both code implementation and computing
hardware properties and b) algorithmic parameters can move cost between these two classes of operation. This means that if
the hardware changes, parameters can be retuned to move algorithmic cost into the operations that are relatively more efficient,
and unlike with standard Krylov solvers we do not typically end up with parameter free black box algorithms.

\begin{figure}[hbt]
\includegraphics*[width=\textwidth]{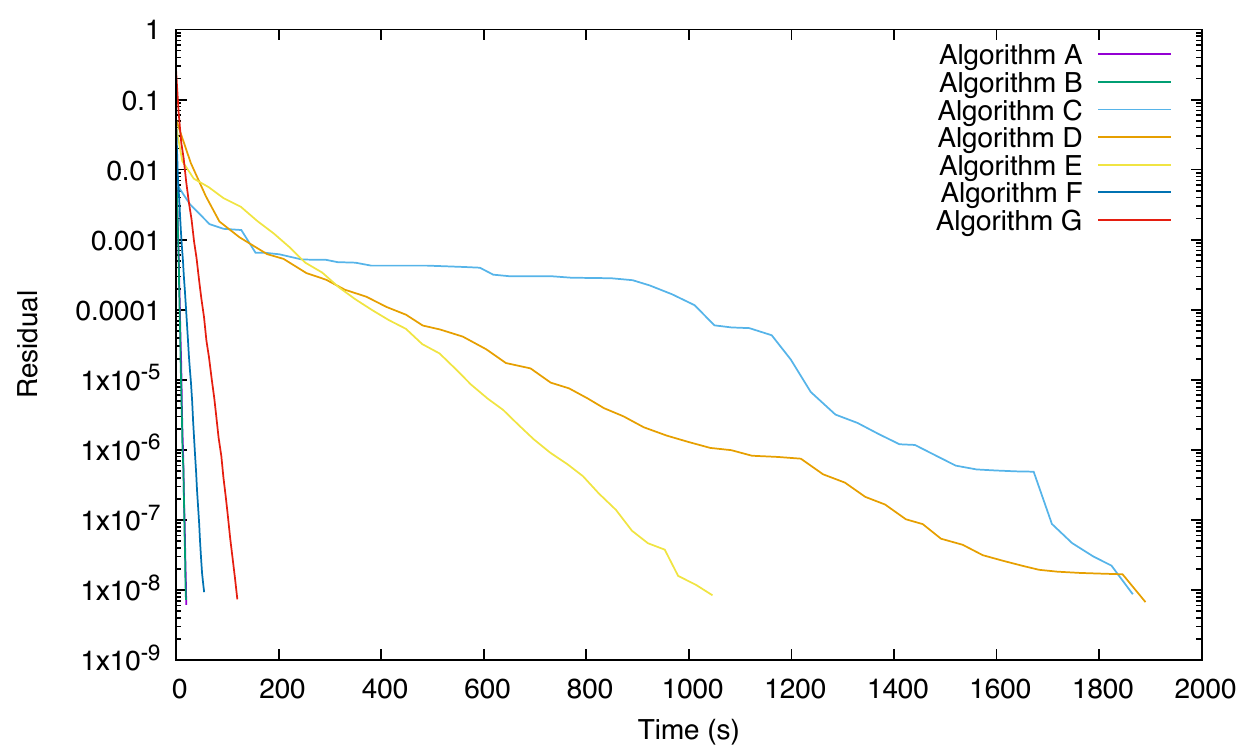}
\caption{\label{fig:Time}
  We plot the convergence history of Algorithms A-G versus time on a single node of the Summit computer.
  This include the entire cost of the algorithm, including coarse grid corrections, and
  is a direct comparison. The software implementation of the coarse Grid solution, while optimised, has not
  necessarily received as much effort as the fine operator. There are some significant caveats to comparing timings
  with undeflated algorithms in the case of Algorithms C, D and badly implemented in the case of Algorithm F. This is explained
  in the body of this text.
}
\end{figure}

For Algorithms C, D, and G the coarse space operations has unrepresentative cost,
because there is no coarse level deflation in Algorithm C and D, and there
is an inefficient but convenient implementation of the coarse operator in Algorithm F.
With this caveat,
Figure~\ref{fig:Time} plots the convergence history of Algorithms A-G versus time on a single node of the Summit
computer. The $16^3$ runs use a single node and make use of four of the six NVIDIA Volta GPU's. The calculation runs over high
bandwidth NVlink for high performance intra-node
communications which perhaps suppresses the overhead of the fine matrix somewhat. The relative behaviour
between multigrid and standard Krylov solvers would differ in a multi-node simulation where interconnect bandwidth
limits would likely affect the fine matrix multiply more, while the coarse grid operator is MPI latency bound.

\begin{figure}[hbt]
\includegraphics*[width=\textwidth]{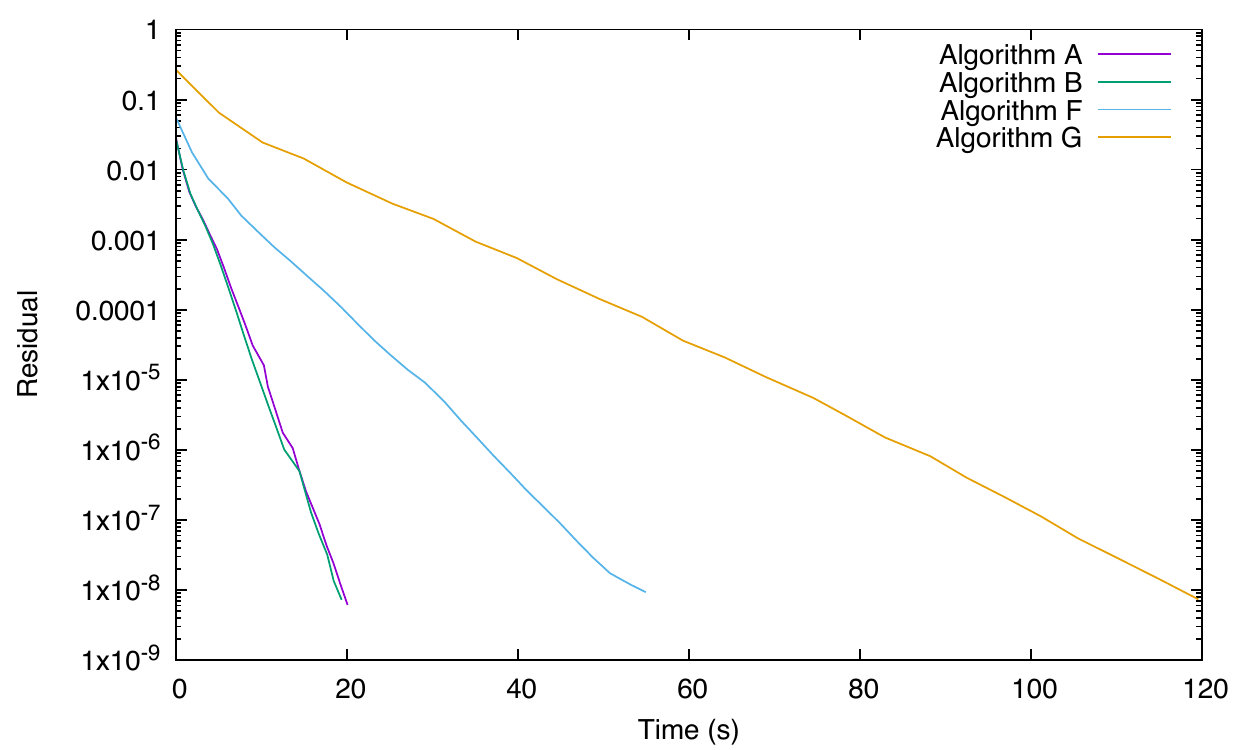}
\caption{\label{fig:TimeFast}
  We plot the convergence history of Algorithms A, B, E, F and G versus time on a single node of the Summit computer.
}
\end{figure}

Figure~\ref{fig:Time} and Figure~\ref{fig:TimeFast} show the convergence of the set of algorithms with and without zoom on the
fastest.
We obtain a three-fold clocktime speed up over CGNR on the red-black preconditioned system and a
nine-fold speed up over the unpreconditioned CGNR algorithm for HDCR (Algorithms A and B).
A modest speed up is obtained for hybrid Algorithm F.

These final plots have a significant caveat for Algorithms C and D,
which suffer from the expense of the five dimensional coarse space, despite showing some deflation effectiveness
when cost is measured solely by fine matrix multiplies. This is partly because a modest blocking factor $2^4$ was used
for the coarsening with $24$ vectors, while the large fifth dimension is retained in the coarse space.
However, we use deflated solves in the coarse
space for Algorithms A,B, E, F and G, but not for Algorithms C and D. We also do not use more than two levels in the multigrid for
Algorithms C and D. We replot the convergence vs time, omitting Algorithms C and D, in figure~\ref{fig:TimeFast}, so that more detailed
comparison of Algorithms A, B, E, F and G may be made.
Algorithm F is a three level extension to Algorithm E, where the fifth dimension is removed in a second step of coarsening.
This doubly coarsened system is then deflated, but the gain is not sufficient to close with either HDCR Algorithms A and B.

The  implementation Algorithm G (HDCG) is very much sub-optimal, since the coarse operator is implemented by simply calling
and projecting the fine operator.
Algorithm G, which used the fewest fine matrix multiplies, had a coarse operator
implemented in an inefficient manner, using the fine matrix and
projecting it, instead of implementing it as a non-local stencil
in the coarse space. Nevertheless, Algorithm G was remarkably competitive due to it using the fewest (compulsory)
fine matrix multiplies. 

\subsubsection{Setup costs}

Minimal setup cost is important to the intended application of these algorithms: the deflation of solutions in the Hybrid
Monte Carlo and related algorithms. In Table~\ref{tab:setup} we display the setup costs for the $16^3$ system with each of the algorithms
studied above, and the relevant timescale for comparison is the total solve time for the red-black preconditioned CGNR
algorithm. In HMC, the setup cost can potentially be amortised across multiple determinant factors and will be the subject of further study.
As one would expect four dimensional coarsenings have a small setup cost, but the cost of the coarse space
 has an additional $L_s$ factor. The area has promise, but we have not been able to make the most obvious approaches give
 a net gain. Overall the most promising option to pursue are variants of the HDCR algorithm, and we will apply both these
 and Algorithm F, Hybrid-$M^\dagger M$ in the next section to larger volumes and (near) physical quark masses.

\begin{table}[hbt]
  \begin{tabular}{
      p{0.14\textwidth}
      p{0.14\textwidth}
      p{0.14\textwidth}
      p{0.18\textwidth}
      p{0.14\textwidth}
      p{0.14\textwidth}}
    Algorithm &
    Coarse\newline Subspace & Coarse\newline operator &
    Eigenvectors &  Lanczos\newline Time & Solve Time \\
    \hline
    rbCGNR    &   -                 &  -              &   -          &   -        & 62s \\
    \hline
    A         &   43s               &  21s            &    48        &  15s  &  20.7s     \\
              &                     &                 &    64        &  20s  &  19.6s     \\
              &                     &                 &   128        &  31s  &  18.6s     \\
    \hline
    A$^\dagger$ &   43s              &  21s            &    128        &  10s  & 23.5s      \\%
    \hline
    B         &   36s               &  14s            &    48        &  29s  &  20.5s    \\
              &                     &                 &   128        &  87s  &  17.7s    \\
    \hline
    B$^\dagger$         &   36s               &  14s            &   128        &  22s  & 18.6s\\
    \hline
    C         &   3.2s              &  1.2s           &    -        &  &   1089s        \\
    \hline
    D         &   3.2s              &  1.2s           &    -        &  &   1554s       \\
    \hline
    E         &   3.2s              &  1.2s           &   -         &     -    & 1051s   \\
    E         &   3.2s              &  1.2s           &    64       &    800s  & 143s  \\
    \hline
    F         &   2.8s + 19s        &  0.8s + 5s        &   128       &   90s    &  54s    \\
    \hline
    G         &   39s               &  -              &    32       &    312s  & 127s   \\
    \hline
  \end{tabular}
  \caption{ \label{tab:setup}
    We display the setup costs for the algorithms studied on the $16^3$ volume.
In algorithms A$^\dagger$ and B$^\dagger$  we removed the requirement that Lanczos converge, and ran the Lanzczos A-orthogonal recurrence relation
for $N_k = N_m = N_{ev}$ vectors and use these as a deflation basis in which the representation of the matrix
is diagonal.
  }
\end{table}

\subsection{Multigrid solver on physical point lattice}

We have reoptimised the most successful of the algorithms studied so far, on a larger system with lighter quark
masses,  with the HDCR parameters tuned in Algorithm H, Table~\ref{tab:algorithmH}, and we also retuned Algorithm F.
We used a single $48^3\times 96$ Mobius domain wall fermion configuration number 1000 from our $m_{ud}=0.00074$ 2+1 flavour Iwasaki gauge
ensemble at $\beta=2.13$\cite{Allton:2007hx} with $L_s=24$. The valence quark action was $L_s=24$ but with the Shamir
Domain Wall action, due to the $\Gamma_5$ hermiticity constraint of the HDCR algorithm. This was therefore non-unitary and had higher than physical
effective quark mass due to the increased residual chiral symmetry breaking. However, this system still serves as a useful test of the algorithm
on larger volumes and lighter quark masses than with our $16^3$ test system.

\begin{table}[hbt]
\begin{tabular}{cc}
  Algorithm & H  \\
  \hline
  Fine   Grid & $48^3\times 96\times 24$ \\
  Block       & $2^4\times 16$ \\
  Coarse Grid & $24^3\times 48\times 1$\\
  \hline
  Outer Krylov  & ${\rm pGCR}(H_{DW})$  \\
  Basis vectors & 40, 32 \\
  Smoother   & $S_{\rm chebyshev}([0.5,60],12)$ \\
  \hline
  Coarsening    & $H_{DW}$  \\
  Coarse Solver & Deflated CGNR  \\
  Coarse Tolerance     & 0.02\\
  Coarse Eigenvectors     & 0 \\
  \hline
  Subspace $\lambda_{\rm max}$ & 60.0  \\
  Subspace $\lambda_{\rm lo}$ & 0.01 \\
  Subspace $m$ & 800  \\
  Subspace $\Delta$ & 100\\
  \hline
\end{tabular}
\caption{ \label{tab:algorithmH} Algorithms H is HDCR\cite{Yamaguchi:2016kop}, applied to a large lattice volume and near physical quark mass.  }
\end{table}

In the larger volume eigenvector deflation of the coarse space was not tractable for an HMC targeted algorithm due
to the greater number of eigenvectors required, while our code had sufficient power of two
constraints that in a 128 node MPI task on the Summit computer, it was also not possible to use a three level
algorithm. Instead we introduced a red-black solver on the coarse space, to give a limited acceleration of convergence.
Table~\ref{tab:physical} displays the timings for both red black preconditioned CGNR and also the setup and solve time for HDCR, both with
and without deflation in the coarse space. It also displays the setup costs.

We can see for Algorithm H, after tuning, we obtained a set up time of 122.5s, and solve time of 149s, and the combined 272s can
be compared to the red-black CGNR solution time of 502s, and unpreconditioned CG of 1612s. For Algorithm F, after tuning, we obtained
a set up time of 62.5s, and solve time of 242s and combined 304.5s. 

This represents a significant net speed up, even including setup costs, and is very encouraging, and perhaps even a breakthrough in
reducing multigrid setup overheads. The fine grid set up costs for Algorithm F are four dimensional, based on the Wilson operator, but
a less efficient solver is obtained.

Since two solves of the unsquared operator are required for the two flavour determinant, but only one for the squared operator, this has passed
the break even point, even without subspace reuse across multiple Hasenbusch determinant ratio factors.
We gain confidence that for the Shamir DWF formulation we can obtain a real acceleration of HMC.
In the Conclusions section~\ref{sec:conclusions} we further discuss the prospects and next steps for moving this into a gain for HMC evolution.

\begin{table}[hbt]
  \begin{tabular}{
      p{0.2\textwidth}
      p{0.2\textwidth}
      p{0.2\textwidth}
      p{0.2\textwidth}
      p{0.2\textwidth}}
    Algorithm &
    Coarse\newline Subspace & Coarse\newline operator &   Solve Time \\
    \hline
      CGNR    &   -                 &  -                                & 1612s \\
    rbCGNR    &   -                 &  -                                & 502s \\
    \hline
    H                        &   110s              &  12.5s                     &  195s    \\
    H (RB CG coarse)         &   110s              &  12.5s                     &  149s    \\
    F (RB CG coarse)         &    50s              &  12.5s                     &  242s    
\end{tabular}
  \caption{ \label{tab:physical} We display the wall clock timing and setup cost for HDCR (Algorithm H) on a physical point ensemble,
    and a new Hybrid-$M^\dagger M$ Algorithm F. The solves have 
  a non-unitary DWF action which was slighly above the physical quark mass due to residual chiral symmetry breaking effects.
  In this system HDCR was able to both setup and solve the linear system with combined time faster than the rbCGNR algorithm.
  This has been enabled by the fast setup algorithms introduced in this work. The use of red-black preconditioning in the coarse space was helpful.
}
\end{table}

\section{Conclusions}
\label{sec:conclusions}

The main findings of this paper are as follows.

{\bf Domain wall fermion multigrid}:
We have compared a number of different schemes for domain wall fermion multigrid algorithms
and find that HDCR\cite{Yamaguchi:2016kop} so far is the most promising direction, with both a significant speed up
over red-black preconditioned conjugage gradients and low setup overhead.

{\bf Fast setup multigrid}:
We have introduced a new scheme for multigrid setup based on spectral filtering
using Chebyshev polynomials. The recursive nature of these polynomials allows multiple
useful vectors to be obtained from a single initial random noise vector.
The method demonstrates substantially reduced setup cost. The aim of this method
is to improve the effectiveness of multigrid algorithms in gauge evolution
algorithms such as the Hybrid Monte Carlo algorithm\cite{Duane:1987de},
the gluon field is changed after a single solution of the Dirac operator (or in the case of
multiple Hasenbusch determinant factors after a modest number of solutions of related operators).

Current state of the art involves polynomial prediction of the deflation basis as the gauge
configuration is evolved, but this prediction both violates reversibility and also
mistracks the evolution of the configuration requiring periodic recalculation or improvement of the
deflation basis vectors. The Metropolis algorithm requires reversibility, so convergence to a
tight stopping criterion is then required. This is tight stopping condition is not required if a reversible
guess (such as a zero guess) is used, and the stopping condition for force evaluation during molecular dynamics
with Krylov solvers steps may be relaxed with reversibility violated only by numerical rounding error, and not
convergence stopping condition. 
One might hope to reduce the setup cost of multigrid algorithms to the point where the setup cost
might be amortised in a single solution, or at very least across several Hasenbsuch determinant ratio factors, whereupon
the convergence tolerance might remain relaxed.

In this work we have achieved a significant step forward, where on our $48^3$ volume we are able to both setup
and solve \emph{twice} the HDCR algorithm significantly faster than the conventional red-black preconditioned CG. The algorithm is restricted
(currently) to the standard Shamir formulation of domain wall fermions, but we believe this is a significant step
towards a genuine speed up for HMC.
We have not, however, obtained the order-of-magnitude solve time gain that multigrid has enabled with the Wilson action.

{\bf Usage in HMC}:
The fermion determinant in domain wall fermions is that of a ratio of the two flavour and Pauli Villars operators,
$$
{\rm det} \frac{M_l^\dagger M_l}{M_{PV}^\dagger M_{PV}},
$$
and this is normally factored as several intermediate Hasenbusch terms such as,
$$
{\rm det} \frac{M(m_l)^\dagger M(m_l)}{M(m_1)^\dagger M(m_1)}
{\rm det} \frac{M(m_1)^\dagger M(m_1)}{M(m_2)^\dagger M(m_2)}
{\rm det} \frac{M(m_2)^\dagger M(m_2)}{M(m_{PV})^\dagger M(m_{PV})}.
$$
The low mode spaces for the denominators in these determinant ratios do not coincide due to the nature of the
domain wall fermion mass term. However these intermediate fragments are unphysical, and alternate forms can be considered.
A frequency splitting scheme that is based on an additive shift is also possible,
$$
{\rm det} \frac{M(m_l)^\dagger M(m_l)}{M(m_l)^\dagger M(m_l)+\Delta_1}
{\rm det} \frac{M(m_l)^\dagger M(m_l)+\Delta_1}{M(m_l)^\dagger M(m_l)+\Delta_2}
{\rm det} \frac{M(m_l)^\dagger M(m_l)+\Delta_2}{M(m_{PV})^\dagger M(m_{PV})},
$$
and may allow share the multigrid setup across multiple inversions.

{\bf Four dimensional coarsening}:
a modest $2^4$ blocking cell and a significant fifth dimension extent was required with four dimensional coarsening
based on the $D_W$ operator. While the deflation provided is in principle effective, the coarse
space cost proves a cost barrier that we were not able to address sufficiently well to establish an
effective method in four dimensional QCD. Comparing to the non-red-black solver, we do obtain a significant 
reduction in fine matrix multiplies from 19,082 to 2744 which is broadly consistent with the D=2 Schwinger model
results\cite{Brower:2020xmc}, but the red-black solver is the more appropriate base line and the cost of the coarse space
is significant. These combine to eliminate the gain.
For the MG-PV algorithm, the BiCGSTAB algorithm as smoother proved to be significantly less effective than the GCR algorithm.

For the MG-$M^\dagger M$ approach, a hybrid coarsening scheme, stepping through one level of Wilson
operator based coarsening and a second level of five dimensional coarsening based on the $H_{DW}$ operator
substantially reduced the cost, and Algorithm F became cheaper than red-black CG on the large volume with light quark mass.
Further study is a good idea, since the setup on the fine space is proportional to the four dimensional
lattice volume, rather than five dimensional system, which intriguingly serves to remove or reduce the extra
cost of chiral fermions. A hybrid scheme has been demonstrated that is close to competitive, and for which
the multigrid setup used is only four dimensional operations on the finest grid.

Our attempts with the non-hermitian MG-PV algorithm have not so far been encouraging, and comparison with the new MG-$M^\dagger M$
and Hybrid-$M^\dagger M$ algorithms suggest that additional effort is required to make these classes of approach
successful. The hybrid scheme, Algorithm F is close to competitive, leaving some encouragement for reducing the poor scaling of
cost with chiral symmetry.

\section{Acknowledgements}

A.Y. has been supported by an Intel Parallel Computing Centre held at the Higgs Centre for Theoretical Physics.
P.B. acknowledges Wolfson Fellowship WM160035, an Alan Turing Fellowship, and 
STFC grants ST/P000630/1, ST/M006530/1, ST/L000458/1, ST/K005790/1, ST/K005804/1, ST/L000458/1.
P.B. has also been supported by DOE contract DESC0012704.
We would like to thank Kate Clark, Evan Weinberg, Dean Howarth and Richard Brower for useful discussions.
We would like to thank Daniel Richtmann and Tilo Wettig for contributions to the 
Grid multigrid code to support Wilson fermions and non-hermitian systems.
We would like to thank Andreas J\"uttner for contributing the BiCGSTAB algorithm code to Grid.
All numerical tests were run on the Summit Supercomputer at Oak Ridge National Laboratory under the
USQCD Exascale Computing Project allocation LGT104.


\begin{thebibliography}{99}

\bibitem{Boyle:2014rwa} 
  P.~A.~Boyle,
  ``Hierarchically deflated conjugate gradient,''
  arXiv:1402.2585 [hep-lat].

\bibitem{Yamaguchi:2016kop}
A.~Yamaguchi and P.~Boyle,
``Hierarchically deflated conjugate residual,''
PoS \textbf{LATTICE2016}, 374 (2016)
doi:10.22323/1.256.0374
[arXiv:1611.06944 [hep-lat]].

\bibitem{Brower:2020xmc}
R.~C.~Brower, M.~A.~Clark, D.~Howarth and E.~S.~Weinberg,
``Multigrid for Chiral Lattice Fermions: Domain Wall,''
[arXiv:2004.07732 [hep-lat]].



\bibitem{Brezina}
M. Brezina, R. Falgout, S. MacLachlan, T. Manteuffel, S. McCormick, and J. Ruge. Adaptive Smoothed Aggregation (aSA) - SIAM J.Sci.Statist.Comput.,25,1896

\bibitem{Luscher:2007se} 
  M.~Luscher,
  ``Local coherence and deflation of the low quark modes in lattice QCD,''
  JHEP {\bf 0707}, 081 (2007)
  doi:10.1088/1126-6708/2007/07/081
  [arXiv:0706.2298 [hep-lat]].

\bibitem{Brannick:2007ue} 
  J.~Brannick, R.~C.~Brower, M.~A.~Clark, J.~C.~Osborn and C.~Rebbi,
  ``Adaptive Multigrid Algorithm for Lattice QCD,''
  Phys.\ Rev.\ Lett.\  {\bf 100}, 041601 (2008)
  doi:10.1103/PhysRevLett.100.041601
  [arXiv:0707.4018 [hep-lat]].

\bibitem{Brannick:2007cc} 
  J.~Brannick, R.~C.~Brower, M.~A.~Clark, J.~C.~Osborn and C.~Rebbi,
  ``Adaptive Multigrid Algorithm for the QCD Dirac-Wilson Operator,''
  PoS LAT {\bf 2007}, 029 (2007)
  [arXiv:0710.3612 [hep-lat]].

\bibitem{Clark:2008nh} 
  M.~A.~Clark, J.~Brannick, R.~C.~Brower, S.~F.~McCormick, T.~A.~Manteuffel, J.~C.~Osborn and C.~Rebbi,
  ``The Removal of critical slowing down,''
  PoS LATTICE {\bf 2008}, 035 (2008)
  [arXiv:0811.4331 [hep-lat]].

\bibitem{Babich:2009pc} 
  R.~Babich, J.~Brannick, R.~C.~Brower, M.~A.~Clark, S.~D.~Cohen, J.~C.~Osborn and C.~Rebbi,
  ``The Role of multigrid algorithms for LQCD,''
  PoS LAT {\bf 2009}, 031 (2009)
  [arXiv:0912.2186 [hep-lat]].

\bibitem{Osborn:2010mb} 
  J.~C.~Osborn, R.~Babich, J.~Brannick, R.~C.~Brower, M.~A.~Clark, S.~D.~Cohen and C.~Rebbi,
  ``Multigrid solver for clover fermions,''
  PoS LATTICE {\bf 2010}, 037 (2010)
  [arXiv:1011.2775 [hep-lat]].

\bibitem{Frommer:2012mv} 
  A.~Frommer, K.~Kahl, S.~Krieg, B.~Leder and M.~Rottmann,
  ``Aggregation-based Multilevel Methods for Lattice QCD,''
  PoS LATTICE {\bf 2011}, 046 (2011)
  [arXiv:1202.2462 [hep-lat]].

\bibitem{Frommer:2013fsa} 
  A.~Frommer, K.~Kahl, S.~Krieg, B.~Leder and M.~Rottmann,
  ``Adaptive Aggregation Based Domain Decomposition Multigrid for the Lattice Wilson Dirac Operator,''
  SIAM J.\ Sci.\ Comput.\  {\bf 36}, A1581 (2014)
  doi:10.1137/130919507
  [arXiv:1303.1377 [hep-lat]].

\bibitem{Frommer:2013kla} 
  A.~Frommer, K.~Kahl, S.~Krieg, B.~Leder and M.~Rottmann,
  ``An adaptive aggregation based domain decomposition multilevel method for the lattice wilson dirac operator: multilevel results,''
  arXiv:1307.6101 [hep-lat].

\bibitem{Weinberg:2017zlv}
E.~S.~Weinberg, R.~C.~Brower, K.~Clark and A.~Strelchenko,
``Progress Report on Staggered Multigrid,''
PoS \textbf{LATTICE2016}, 273 (2017)
doi:10.22323/1.256.0273

\bibitem{Brower:2018ymy}
R.~C.~Brower, M.~A.~Clark, A.~Strelchenko and E.~Weinberg,
``Multigrid algorithm for staggered lattice fermions,''
Phys. Rev. D \textbf{97}, no.11, 114513 (2018)
doi:10.1103/PhysRevD.97.114513
[arXiv:1801.07823 [hep-lat]].
  
\bibitem{Richtmann:2019eyj}
D.~Richtmann, P.~A.~Boyle and T.~Wettig,
``Multigrid for Wilson Clover Fermions in Grid,''
PoS \textbf{LATTICE2018}, 032 (2019)
doi:10.22323/1.334.0032
[arXiv:1904.08678 [hep-lat]].

\bibitem{Kaplan:1992bt} 
  D.~B.~Kaplan,
  ``A Method for simulating chiral fermions on the lattice,''
  Phys.\ Lett.\ B {\bf 288}, 342 (1992)
  doi:10.1016/0370-2693(92)91112-M
  [hep-lat/9206013].

\bibitem{Shamir:1993zy} 
  Y.~Shamir,
  ``Chiral fermions from lattice boundaries,''
  Nucl.\ Phys.\ B {\bf 406}, 90 (1993)
  doi:10.1016/0550-3213(93)90162-I
  [hep-lat/9303005].
\bibitem{Cohen:2012sh} 
  S.~D.~Cohen, R.~C.~Brower, M.~A.~Clark and J.~C.~Osborn,
  ``Multigrid Algorithms for Domain-Wall Fermions,''
  PoS LATTICE {\bf 2011}, 030 (2011)
  [arXiv:1205.2933 [hep-lat]].

  \bibitem{Clark:2017wom}
M.~A.~Clark, C.~Jung and C.~Lehner,
``Multi-Grid Lanczos,''
EPJ Web Conf. \textbf{175}, 14023 (2018)
doi:10.1051/epjconf/201817514023
[arXiv:1710.06884 [hep-lat]].

\bibitem{Babich:2010qb}
R.~Babich, J.~Brannick, R.~C.~Brower, M.~A.~Clark, T.~A.~Manteuffel, S.~F.~McCormick, J.~C.~Osborn and C.~Rebbi,
``Adaptive multigrid algorithm for the lattice Wilson-Dirac operator,''
Phys. Rev. Lett. \textbf{105}, 201602 (2010)
doi:10.1103/PhysRevLett.105.201602
[arXiv:1005.3043 [hep-lat]].


\bibitem{trefethen} 
N. M. Nachtigal, S. C. Reddy, and L. N. Trefethen, ''How Fast are Nonsymmetric Matrix Iterations?'' SIAM. J. Matrix Anal. Appl., 13(3), 778795.


\bibitem{Kennedy:2006ax} 
  A.~D.~Kennedy,
  ``Algorithms for dynamical fermions,''
  hep-lat/0607038.

\bibitem{rudy} 
Rudy Arthur, PhD thesis, University of Edinburgh, 2012.

\bibitem{Saad}
  SIAM: Iterative Methods for Sparse Linear Systems
  Yousef Saad. ISBN: 978-0-89871-534-7
  https://doi.org/10.1137/1.9780898718003

\bibitem{Boyle:2016lbp} 
  P.~A.~Boyle, G.~Cossu, A.~Yamaguchi and A.~Portelli,
  ``Grid: A next generation data parallel C++ QCD library,''
  PoS LATTICE {\bf 2015}, 023 (2016).

\bibitem{Aoki:1997xg}
S.~Aoki and Y.~Taniguchi,
``One loop calculation in lattice QCD with domain wall quarks,''
Phys. Rev. D \textbf{59}, 054510 (1999)
doi:10.1103/PhysRevD.59.054510
[arXiv:hep-lat/9711004 [hep-lat]].

\bibitem{Allton:2007hx}
C.~Allton \textit{et al.} [RBC and UKQCD],
``2+1 flavor domain wall QCD on a (2 fm)*83 lattice: Light meson spectroscopy with L(s) = 16,''
Phys. Rev. D \textbf{76}, 014504 (2007)
doi:10.1103/PhysRevD.76.014504
[arXiv:hep-lat/0701013 [hep-lat]].

\bibitem{CGNR}
  Hestenes, Magnus R.; Stiefel, Eduard.
  ``Methods of Conjugate Gradients for Solving Linear Systems''.
  Journal of Research of the National Bureau of Standards. 49 (6): 409. 1952.

\bibitem{GCR}
  S. C. Eisenstat, H. C. Elman, and H. C. Schultz.
  ``Variational iterative methods for nonsymmetric systems of linear equations.''
  SIAM J. Numer. Anal., 20, 1983  

\bibitem{BiCGSTAB}
H. A. van der Vorst
SIAM J. Sci. and Stat. Comput., 13(2), 631–644. (14 pages)
``Bi-CGSTAB: A Fast and Smoothly Converging Variant of Bi-CG for the Solution of Nonsymmetric Linear Systems''

\bibitem{Duane:1987de}
S.~Duane, A.~D.~Kennedy, B.~J.~Pendleton and D.~Roweth,
``Hybrid Monte Carlo,''
Phys. Lett. B \textbf{195} (1987), 216-222
doi:10.1016/0370-2693(87)91197-X

\bibitem{Kate} Suggested to the authors by Kate Clark.
  
\end{thebibliography}
\end{document}